\documentclass[11pt]{article}
\usepackage[cp1251]{inputenc}
\usepackage[T2A]{fontenc}
\usepackage[english]{babel}
\usepackage{amsfonts}
\usepackage{amsmath}
\usepackage{amssymb}
\usepackage{graphicx}

\newcommand\eps{\varepsilon}
\newcommand\R{{\mathbb R}}

\newtheorem{theorem}{Theorem}

\newtheorem{remark}{Remark}

\title{Solitons and cavitons in a nonlocal Whitham equation}
\author{N. Kulagin$^1$, L. Lerman$^{2,3}$, A. Malkin$^1$, \\
\normalsize
$^1$Institute of Physical Chemistry and Electrochemistry,\\
\normalsize
Russian Academy of Science, Moscow, Russia,\\
\normalsize
$^2$Dept. of Fund. Math. of Higher School of Economics, and\\
\normalsize
$^3$Lobachevsky State University of Nizhny Novgorod, Russia}
\date{}
\begin{document}
\maketitle

\begin{abstract}
Solitons and cavitons (localized solutions with singularities) for the
nonlocal Whitham equations are studied. The equation of a fourth order
with a parameter in front of fourth derivative for traveling waves is
reduced to a reversible Hamiltonian system defined
on a two-sheeted four-dimensional space. When this parameter is small we get
a slow-fast Hamiltonian system. Solutions of the system which
stay on one sheet represent smooth solutions of the equation but those which perform
transitions through the branching plane represent solutions with jumps.
They correspond to solutions with singularities -- breaks of the first and
third derivatives but continuous even derivatives.
The system has two types of equilibria on different sheets, they can of
saddle-center or saddle-foci. Using analytic and numerical methods we
found many types of homoclinic (and periodic as well) orbits to these
equilibria both with a monotone asymptotics and oscillating ones. They correspond
to solitons and cavitons of the initial equation. When we deal with homoclinic
orbits to a saddle-center the values of the second parameter (physical wave speed)
is discrete but for the case of a saddle-center it is continuous.
The presence of majority such solutions displays the very complicated dynamics
of the system.
\end{abstract}

\section{Introduction}

Nonlinear nonlocal Whitham equation
\begin{equation}\label{Eq-1}
  \frac{\partial V}{\partial t}+V\frac{\partial V}{\partial x}=
  \frac{\partial}{\partial x}\int\limits_{-\infty}^{\infty}dx^\prime R(x-x^\prime)
V(x^\prime,t)
\end{equation}
represents a wide class of equations which are of great interest
for nonlinear wave theory. It combines the typical hydrodynamic nonlinearity and
an integral term descriptive of dispersion of the linear theory.
The kernel of the integral term is conventionally defined by the
dispersion relation $\omega = k\widetilde{R}(k)$ with
\begin{equation}\label{Eq-2}
  R(x)=\int\limits_{-\infty}^{\infty}\widetilde{R}(k)e^{-ikx}dk .
\end{equation}
Eq.(\ref{Eq-1}) with $\widetilde{R}= (1+k^2)^{-1}$ was proposed by G.
Whitham instead of Korteweg-de Vries equation in order to describe sharp
crests of the water waves of a greatest height \cite{Whi}.

The usage of relatively simple Whitham type equations appeared to be very
fruitful for various physical applications. A number of special cases of
Eq.(\ref{Eq-1}) were examined in detail.
Among them are the Benjamin-Ono \cite{Be,Case} and Joseph \cite{Joseph} equations
describing internal waves in stratified fluids of infinite and finite depth.
These equations appeared to be integrable by the inverse scattering technique
and the behavior of their solutions has been studied rather well.
The Benjamin-Ono and Joseph equations are however the only representatives
of the Whitham equations possessing this property \cite{Cite-1}.
Another widely known equations of that class were studied not so exhaustively,
although the literature on the subject is quite extensive. A list of well-known
Whitham equations involves the Leibovitz one for the waves
in rotating fluid \cite{Leib}, the Klimontovich equation for magnetohydrodynamic
waves in non-isothermal collision-less plasma \cite{Klim}, equations for shallow
water waves \cite{Whi}, capillary \cite{AK} and hydroelastic \cite{DKMP} waves.
The review on nonlinear nonlocal equations in theory of waves is presented in
detailed monograph \cite{Cite-2}.

The characteristic feature of conservative Whitham equations is the
existence of solitary wave solutions. For all just listed equations these
solutions are smooth except some limiting cases of peaking for the waves of
greatest amplitude. Besides, the amplitudes and velocity spectra of solitons
can be bounded or not. But in any case the spectra are continuous. These
properties are believed to be typical, but, as will be shown below, they are not
essential for the solitons of Whitham equations.

Here we examine a particular case of the Whitham equation with a
resonance dispersion relation
\begin{equation}\label{Eq-3}
  \widetilde{R}(k)=\frac{1}{1-k^2+D^2 k^4}.
\end{equation}
That equation has been proposed for nonlinear acoustic waves in simple peristaltic
systems \cite{Malkin}. With small $D^2$ it is also applicable to the waves in a medium
with internal oscillators \cite{Cite-4}. A tentative analysis of some peculiarities
of solitons to that equation has been performed in a short communication \cite{PRL}.

We study here specific features of solitary wave solutions to
Eqs.(\ref{Eq-1})-(\ref{Eq-3}). It is shown that this equation possesses both
smooth and singularity involving solitons with exponential asymptotics,
bound states of solitons and solitary waves with oscillating asymptotics.
The velocity spectra of exponentially localized solitons turn out to be
discrete ones.

\section{Equation for traveling waves and its reduction}

Hereafter we shall study an ordinary differential equation that obtained by
the inversion of integral operator defined by Eqs (\ref{Eq-1})-(\ref{Eq-3})
and transferring to the traveling wave solutions. The result takes the form of
the fourth order differential equation
\begin{equation}\label{4ord}
D^2 S^{(IV)} + S^{\prime\prime} + S = V,\;S = \lambda V + \frac{1}{2}V^2,
\end{equation}
with traveling coordinate $y= x+\lambda t$, boundary conditions $\lim S(y) = 0$,
as $|y|\to\infty$, and parameters $D^2 << 1,$ $0\le \lambda \le 1$. Thus, physically
treated, the problem of searching for solutions of this type can be thought as
a nonlinear boundary value problem for the parameter $\lambda.$

This equation for very small $D^2$ is of the singularly perturbed type similar
to many such equations, see, for instance, \cite{AK,HM,AM,AK,IK,Segur,GL,Champ}.
Another feature of this equation is its relation with the class of
implicit differential equations \cite{Ar_add}. To see this, let us introduce,
similar to \cite{KLSh}, new variables $u_1=S,\; u_2 = S',$ $v_1 = -S' -
D^2 S^{\prime\prime\prime},$ $v_2 = DS^{\prime\prime}$. Then the equation is reduced
to the Hamiltonian system
\begin{equation}\label{ini}
u_1' = u_2,\;v_1' = u_1- V(u_1),\;Du_2' = v_2,\; Dv_2' = -u_2 - v_1,
\end{equation}
where $V(u_1)$ is given by solving the equation $2u_1 = 2\lambda V + V^2$ w.r.t. $V$.
This  system is also reversible w.r.t. the involution $L: (u_1, v_1,u_2,v_2)\to
$ $(u_1, -v_1,-u_2,v_2),$ i.e. if $(u_1(y), v_1(y),u_2(y),v_2(y))$ is a solution
to the system, then $(u_1(-y), -v_1(-y),-u_2(-y),v_2(-y))$ is as well
\cite{Dev1}. The fixed point set $Fix(L)$ of $L$ is 2-plane $v_1=u_2 =0.$

The quadratic equation for $V(u_1)$ has generally either two or no real solutions, so
function $V$ is two-valued. To keep this into account in more convenient way, let us
consider the space $\R^5$ with coordinates $(u_1,v_1,u_2,v_2,V)$
and its  smooth 4-dimensional submanifold $M$ given by real solutions of the equation
$V^2 + 2\lambda V - 2 u_1 =0.$ This is a two-sheeted submanifold with respect to the
projection $\pi:(u_1,v_1,u_2,v_2,V)\to (u_1,v_1,u_2,v_2)$, its image is half-space
$u_1\ge - \lambda^2/2$. Both sheets are glued along 3-dimensional branching 3-plane
$u_1 = -\lambda^2/2,$ $V = -\lambda.$ The shape of this submanifold is the direct product
of a parabola and a 3-plane.

On each sheet (upper one $V(u_1) = -\lambda + \sqrt{\lambda^2 + 2u_1}$ and lower one
$V = -\lambda - \sqrt{\lambda^2 + 2u_1}$) the system generates its own differential
system. In the half-space $u_1\ge - \lambda^2/2$ on every sheet the Peano theorem on the
existence of solutions is valid \cite{Hartman}, but in the open half-space
$u_1 > - \lambda^2/2$ the usual theorem of existence and uniqueness of solutions works,
so, despite of the non-smoothness of the function $\sqrt{\lambda^2 + 2u_1}$ on the
boundary, through every point in the closed sheet of the open half-space an orbit
passes and the only orbit through the point on a sheet over the open half-space.
There are two questions here: 1) how does one need to adjoin orbits from different sheets
in order to preserve continuity of $S(y)$, when the related orbits hit the boundary
of a sheet, i.e. they satisfy the equality $u_1(y_0) = -\lambda^2/2$ and for
this $y_0$ we have $u_2(y_0)\ne 0$, and 2) about possible non-uniqueness for orbits
through boundary points where $u_1 = - \lambda^2/2$ (and $V= -\lambda$).

The inequality $u_2(y_0)\ne 0$ means that the related orbit must leave a sheet
or enter to a sheet, in dependence on the sign $u_2(y_0)$. Indeed, the restrictions
of both vector fields to the branching plane $u_1 = -\lambda^2/2$ coincide. Hence,
if $u_2(y_0)> 0$ at the boundary point, then the orbit looks inward the region
$u_1 > -\lambda^2/2$ on both sheets and should enter to both of them (we remind
$u'_1 = u_2$). But it is impossible, if $u_1(y)$ (i.e. $S(y)$) varies continuously
in $y$ near this value $y_0$.

If, on the contrary, one gets $u_2(y_0)< 0$, then the orbits look outward in
both sheets and should leave an either sheet. Thus, in order to preserve
continuity of $u_1$ we need to use the reversibility of vector fields and
make jumps on the boundary of both sheets in accordance to the action of $L.$
How does this save the situation, one can see as follows. Suppose
$u_2(y_0)>0$ at the boundary point $X_0 = (u_1(y_0),v_1(y_0),u_2(y_0),v_2(y_0))$,
$u_1(y_0) = -\lambda^2/2$, the vector field looks inward. For $u_2 \ne 0$, the
boundary point does not belong to the set $Fix(L)$ and its $L$-image is another
boundary point $X_0^{(1)}=L(X_0)$ with coordinates $(u_1(y_0),-v_1(y_0),$
$-u_2(y_0),v_2(y_0)).$ Thus, the vector field at $X_0^{(1)}$ looks outward the region
$u_1 > -\lambda^2/2.$ So, suppose we move for $y < y_0$, as $y$ increases to $y_0$,
in the lower sheet along the orbit which hits the branching plane at the point
$X_0^{(1)}$. Next we jump to the point $L(X_0^{(1)})=X_0$ and move further in the
upper sheet along the orbit through the point $X_0$. In this motion the value of $u_1=S$
varies continuously, but $u_2= S'$ undergoes a jump at $y_=y_0.$ There is a
pairing composed orbit, where we move first for $y<y_0$ on the upper sheet till the
point $X_0^{(1)}$, then jump to $X_0$ and go further on the lower sheet along
the orbit through $X_0$. Observe that if an orbit in a sheet does not
cross the branching point its behavior is defined by the smooth (in fact --
analytic) vector field and any tool working in this case can be used. Our
main concern below will be on solutions to (\ref{ini}) which are
homoclinic orbits to equilibria that exist in the system. As we shall see,
these solutions are symmetric but they can be either smooth or with
singularities (they cross the branching plane several times). Smooth
homoclinic orbits we call sometimes solitons and those with singularities
cavitons. Here we follow our terminology in \cite{PRL}.

In order to facilitate simulations, we can eliminate jumps as follows.
The submanifold $M$ is the graph of a smooth function $u_1 = \lambda V + V^2/2$
in variables $(V,v_1,u_2,v_2)$.  Let us rewrite the system in variables
$(r,v_1,u_2,v_2)$, $r=V+\lambda$
$$
rr' = u_2,\;v_1'= \lambda(1-\lambda/2) - r + r^2/2,\;
Du_2'=v_2, Dv_2'=-(u_2+v_1).
$$
The system obtained has singularities along 3-plane $r=0$ (it is not defined).
For upper sheet we get $r = V+\lambda = \sqrt{\lambda^2 + 2u_1}> 0$, but for lower
sheet the sign is opposite $r = V+\lambda = -\sqrt{\lambda^2 + 2u_1} < 0$. In order
to eliminate the singularity, we multiply equations 2-4 at $r$ and change the ``time''
to $s,$ $ds=dy/r,$ obtaining a smooth differential system. The orbits
and direction of moving along orbits of this smooth system coincide on the upper
sheet $r > 0$ with those for orbits of the initial system (\ref{ini}), but on the
lower sheet $r < 0$ the true direction of moving along orbits is opposite. In particular, this
approach allows one to assert that through any point of the boundary $r=0$,
if this point is not singular, a unique orbit can pass on either sheet. This smooth
system looks as follows
\begin{equation}\label{smooth}
\frac{dr}{ds} = u_2,\;\frac{dv_1}{ds} = \lambda(1-\lambda/2)r - r^2
+ r^3/2,\;D\frac{du_2}{ds}= rv_2, D\frac{dv_2}{ds}=-r(u_2+v_1).
\end{equation}
Additional equilibria of the system, appeared due to the change of ``time'', fill
the plane $r=0$, $u_2=0.$

The system (\ref{smooth}) is Hamiltonian and reversible w.r.t. the involution
$L:(r,v_1,u_2,v_2)\to (r,-v_1,-u_2,v_2)$ with the smooth Hamiltonian
$$
H= u_2v_1 + \frac{1}{2}(u_2^2 + v_2^2) - \lambda(1 - \lambda/2)r^2/2 + \frac{1}{3}r^{3}
- \frac{1}{8}r^4.
$$
The symplectic structure here is not standard and defined by the matrix $J$
$$
J = \begin{pmatrix}
0&1&0&0\\-1&0&0&0\\0&0&0&rD^{-1}\\0&0&-rD^{-1}&0
\end{pmatrix},\quad \dot X = J \nabla H.
$$
It is seen that this structure degenerates at $r=0.$

\begin{figure}[!ht]
\begin{minipage}[ht]{0.4\linewidth}
\center{\includegraphics[width=1\linewidth]{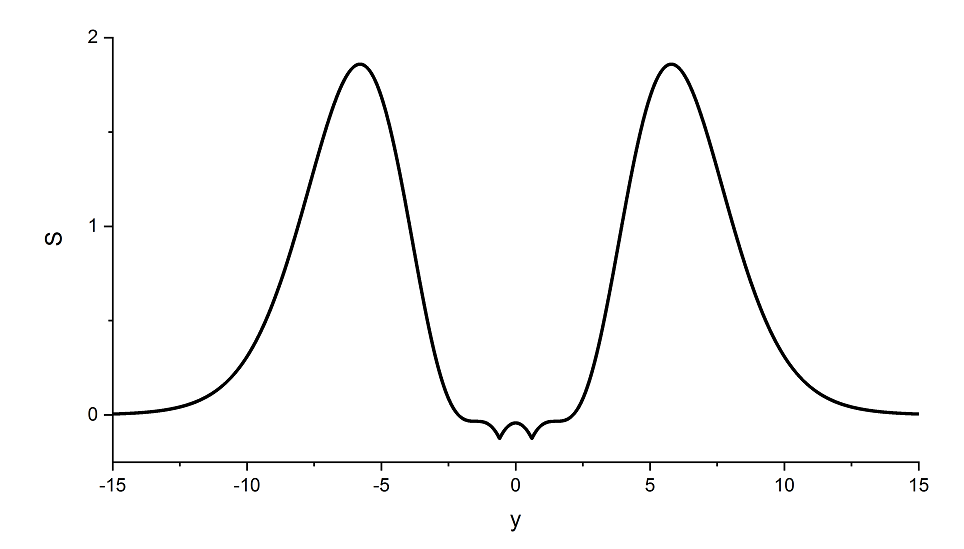} \\ a) }
\end{minipage}
\hfill
\begin{minipage}[ht]{0.4\linewidth}
\center{\includegraphics[width=1\linewidth]{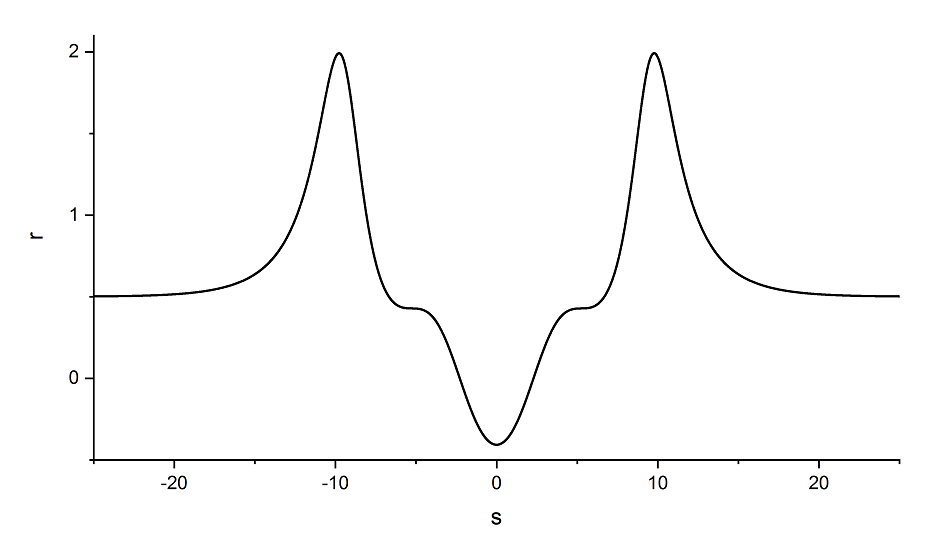} \\ b) }
\end{minipage}
\caption{{\footnotesize (a) Graph of a true 2-round soliton-caviton $S(y)$; (b)
and its smoothing $r(s)$.}}
\label{fig:1}
\end{figure}

The transition to the smooth system sets the question: what are
interrelations between orbits of the smooth system (\ref{smooth}) and
orbits of the initial system on the upper and lower sheets? The answer is
the same as was done above: for orbits not crossing the plane $r=0$ the
behavior is the same as for its counterpart. For $r>0$ direction in $y$
and direction in $s$ are the same, for $r<0$ orbits are the same but
direction of motion is opposite. Situation for orbits crossing $r=0$
appears as before: if an orbit from upper sheet ($r>0$) hits the plane
$r=0$ at a point $X_0$ the true motion is described as above. We make jump
by the action of $L$, $X_1 = L(X_0)$ and after that we continue by the orbit
through $X_1$ in the lower sheet ($r<0$). If we start from the lower
sheet, the procedure is similar.
At such transformation the function $S$ varies
continuously, as well as its $S_{yy}, S_{yyyy}$ but $S_y$ and $S_{yyy}$
change their signs. This leads to the solutions with a singularities (sharpening).
The shape of a dependence $S(y)$, its smoothen variant is plotted on
Fig.~\ref{fig:1}a,b and the projection on $(r,u_2)$-plane is on Fig.~\ref{fig:2}.
The explanations on the orbit behavior will be given below.
\begin{remark}
It is worth remarking that when we deal with symmetric orbits (invariant w.r.t.
the action of $L$), even though they cross the plane $r=0$, the orbits (curves
in the phase space) with be the same as for related orbits of the smooth
system, since, due to symmetry, the passage to the symmetric point occurs
on the same orbit. We use this under simulations.
\end{remark}

\begin{figure}[ht]
	\centering
	\includegraphics[width=0.5\linewidth]{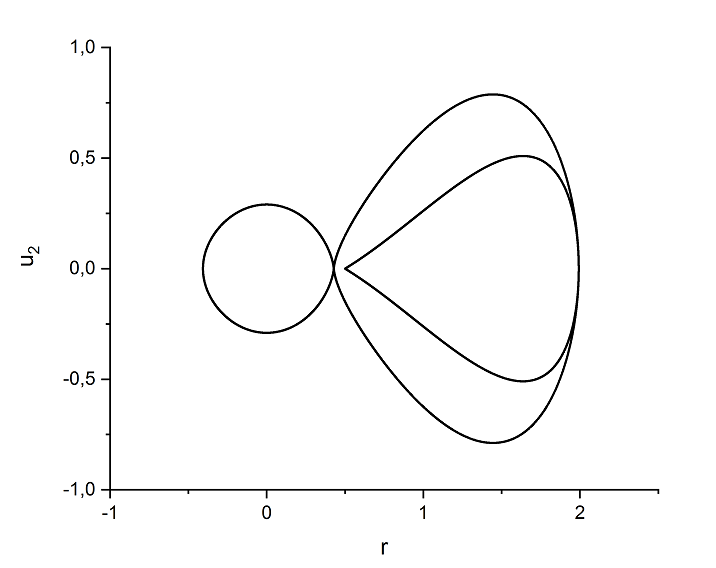}
	\caption{Projection of smoothen soliton-caviton on $(u_2,r)$-plane}
\label{fig:2}
\end{figure}

\section{Slow system}

Let us demonstrate this approach for the limiting system as $D = 0$ (slow system
\cite{Ar_add}).
Then third and fourth equation give the representation for the so-called slow
manifold $v_2 =0, u_2 = -v_1.$ Inserting them into the first and second equations
we get the slow system in the half-plane $u_1\ge -\lambda^2/2$
\begin{equation}\label{abr}
u_1' = -v_1,\;v_1' = u_1 +\lambda \mp \sqrt{\lambda^2 + 2u_1},
\end{equation}
where upper sign corresponds to the upper system and lower sign does to
the lower system.
First we investigate the systems on the upper and lower sheets separately
and compare their behavior with the smooth system obtained by the change
of ``time'' $y.$

\begin{figure}[ht]
	\centering
	\includegraphics[width=0.6\linewidth]{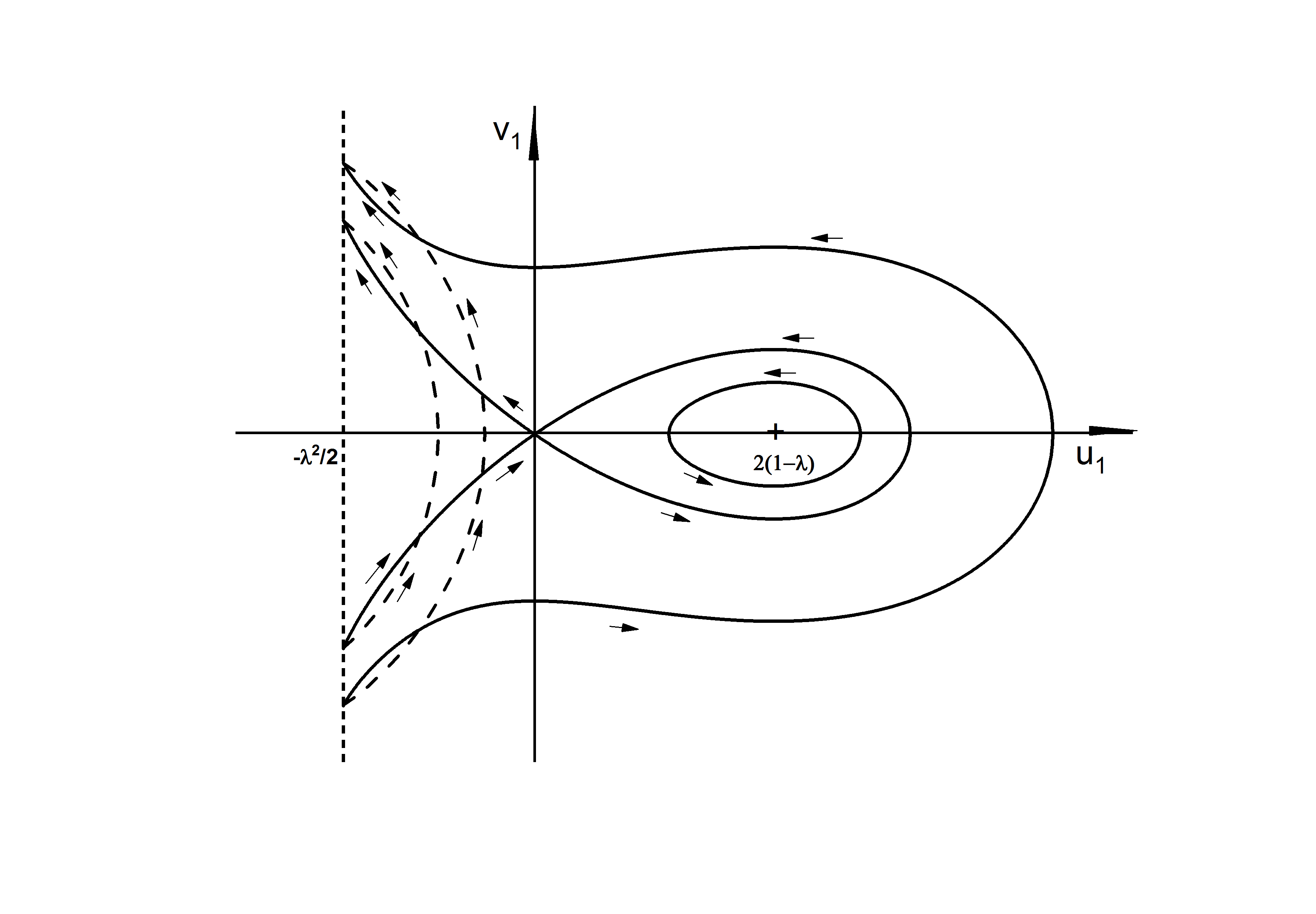}
	\caption{Slow manifold dynamics with jumps.}
\label{fig:slow_break}
\end{figure}

Both systems are Hamiltonian ones and reversible w.r.t. the symmetry $(u_1,v_1)$
$\to (u_1, -v_1).$ The upper system has two equilibria $(0,0)$ and
$2(1-\lambda),$ being a saddle and a center. The lower system has not
equilibria in the half-plane $u_1\ge -\lambda^2/2$ and all their orbits go
from the points $u_1 = -\lambda^2/2$, $|v_1| < \infty$ from negative $v_1$
to hit this line at points with positive $v_1.$ An orbit on the upper sheet
hitting, as $y$ increases, the line $u_1 = -\lambda^2/2$ at the point $(-\lambda^2/2,v_1^*)$,
has to be continued from the point $(-\lambda^2/2,-v_1^*)$ with the further increasing
$y$. The similar is done for orbits on the lower sheet. Under this
procedure we get either periodic orbits (smooth, if it belongs to the upper sheet,
or with a sharpening, if this orbit intersects the line $u_1 =
-\lambda^2/2$). The phase portraits are the following (see, Fig.~\ref{fig:slow_break})
where solid lined represent orbits of the upper system and dashed line do
those for the lower system.

Now let us go to the smooth system in variables $(r,v_1),$ here $r$ can
take any sign.
\begin{equation}\label{abrsl}
r' = -v_1,\;v_1'=\lambda(1-\lambda/2)r - r^2 + r^3/2.
\end{equation}
The Hamiltonian of the system is
$$
h= \frac{v_1^2}{2} + \frac{\lambda(2-\lambda)r^2}{4} - \frac{r^3}{3} + \frac{r^4}{8}.
$$
The system is also reversible with respect
to the involution $L:(r,v_1)\to (r, -v_1)$. This implies that if $(r(s),v_1(s))$ is a
solution, then $(r(-s),-v_1(-s))$ as well.

\begin{figure}[ht]
	\centering
	\includegraphics[width=0.5\linewidth]{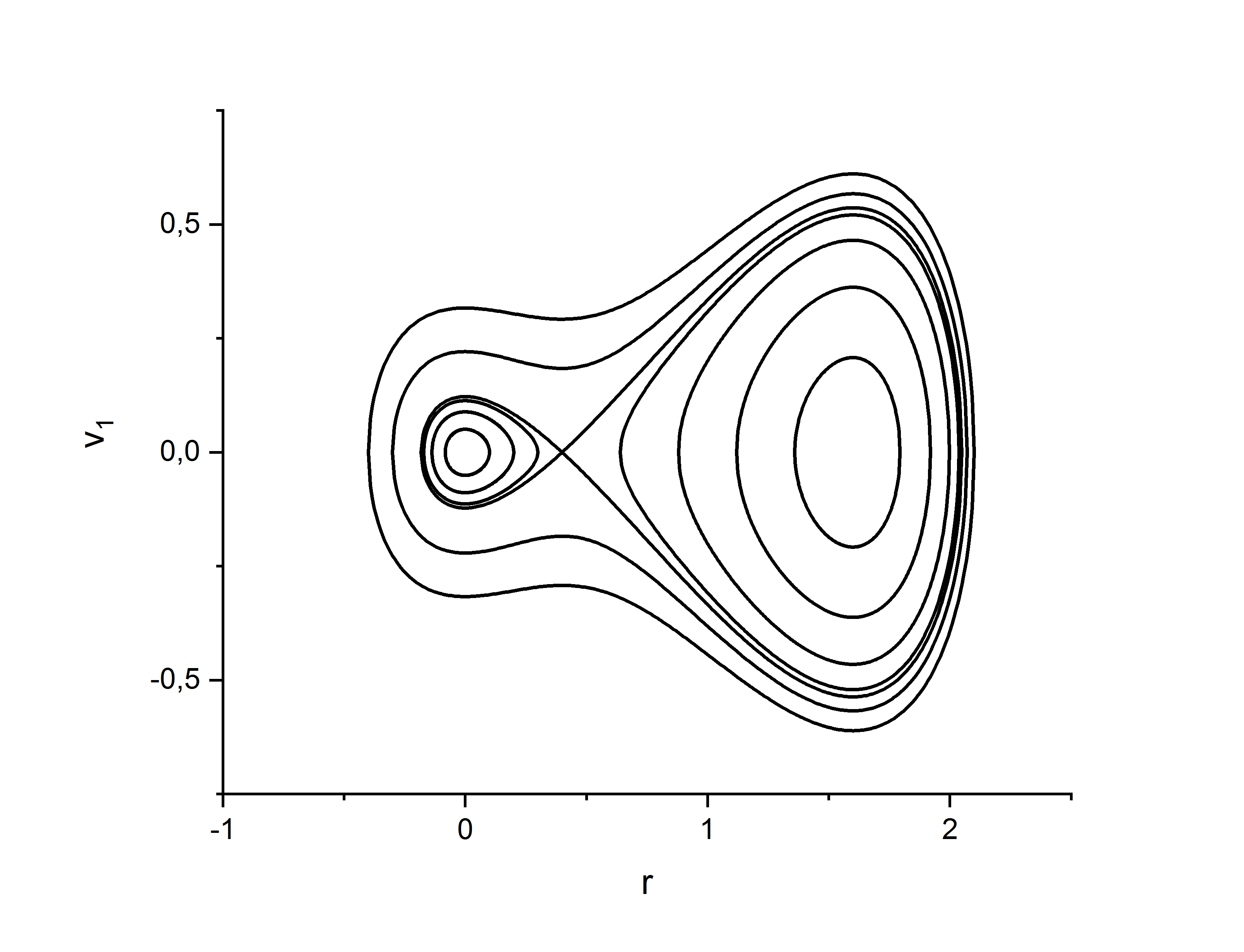}
	\caption{Phase portrait for smoothen system (\ref{abrsl}).}
\label{fig:smooth_ss}
\end{figure}

The equilibria of the system are $v_1=0, r = 0, \lambda, 2-\lambda$, of
them the first and third are centers, the second is a saddle with two separatrix
loops. The left center corresponds to the point of the glued system on the
line $u_1= -\lambda^2/2$ where $v_1=0$ (the point of tangency for orbits of the line).
Here all orbits, except for two loops and equilibria, are periodic ones.
Thus we get the following plot (Fig.~\ref{fig:smooth_ss}).

\section{$\lambda = 0$}

Here we consider first the degenerated limiting case $\lambda =0$ (in fact, physically
not relevant).
This consideration is useful to compare with results for values $\lambda > 0.$
Thus, we get two systems defined in the half-space $u_1\ge 0$, on the upper sheet
\begin{equation}\label{upper0}
u_1' = u_2,\;v_1' = u_1 - \sqrt{2u_1},\;Du_2' = v_2,\; Dv_2' = - u_2 - v_1,
\end{equation}
with equilibria $(0,0,0,0)$ and $(2,0,0,0)$
and on the lower sheet
\begin{equation}\label{lower0}
u_1' = u_2,\;v_1' = u_1+ \sqrt{2u_1},\;Du_2' = v_2,\; Dv_2' = -u_2 - v_1
\end{equation}
with the only equilibrium $(0,0,0,0).$

Smoothing the system in variables $(r,v_1,u_2,v_2)$ with changing time
gives the system ($r$ is arbitrary)
\begin{equation}\label{fin0}
r' = u_2,\;v_1'= - r^2 + r^3/2,\;Du_2'=rv_2, Dv_2'=-r(u_2+v_1).
\end{equation}
with integral $H_0 = u_2v_1 + (u_2^2+v_2^2)/2 + r^3/3 - r^4/8.$ Also the system
is reversible w.r.t. the involution $L: (r,v_1,u_2,v_2)\to (r,-v_1,-u_2,v_2)$, its
fixed points plane is given as $v_1=u_2 =0.$

The system has a plane $P_0$ of equilibria $r=u_2=0.$ Only a curve
from this plane belongs to the level $H=c$. In particular, in the level
$H=0$ where orbits asymptotic to equilibrium at the origin
$r=v_1=u_2=v_2=0$ lie, the intersection of $P_0$ with this level is given
in coordinates $v_1,v_2$ as straight line $(v_1^0,0).$

\begin{figure}[!ht]
\begin{minipage}[ht]{0.4\linewidth}
\center{\includegraphics[width=1\linewidth]{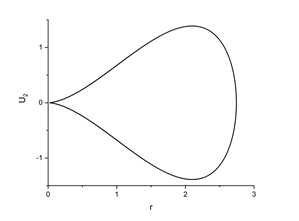} \\ a) }
\end{minipage}
\hfill
\begin{minipage}[ht]{0.4\linewidth}
\center{\includegraphics[width=1\linewidth]{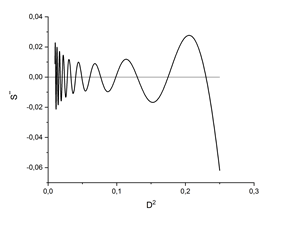} \\ b) }
\end{minipage}
\caption{{\footnotesize (a) The projection of the homoclinic orbit
on the plane $(r,u_2)$; (b)
Graph $S^{\prime\prime\prime}(D^2)$.}}
\label{fig:5}
\end{figure}

The system obtained has a complex equilibrium $O$ at the origin: all four its
eigenvalues are zeroth. Hence the study of its local orbit behavior is a
rather complicated problem. In the first turn we are interested in its
orbits that enter and leave the equilibrium $O$ as $s\to \pm \infty.$ Such
orbits, if they exist, have power asymptotics in $s\to \infty .$ Let us find these
asymptotics using the following ansatz
\begin{equation}
r = As^{-\alpha}(1+o(1)),\;v_1 = Bs^{-\beta}(1+o(1)),\;u_2 = Cs^{-\gamma}(1+o(1)),\;v_2 =
Es^{-\delta}(1+o(1)),
\end{equation}
with unknown coefficients $A,B,C,E$ and exponents $\alpha, \beta, \gamma,
\delta$ to be found. Inserting these functions into differential equations
(\ref{fin0}) we get exponents $\alpha = 4/3,\;\beta = 5/3,\;\gamma = 7/3,\;\delta =
2=6/3$ and coefficients
\begin{equation}\label{expo}
A=\frac{2\sqrt[3]{35}}{3}D^{2/3},\;
B=\frac{4}{15}\sqrt[3]{35^2}D^{4/3},\; C=-\frac{8\sqrt[3]{35}}{9}D^{2/3},\;
E=\frac{28}{9}D.
\end{equation}
The same calculation with the change $s = -s_1$ to get orbits tending $O$
as $s\to -\infty$ gives naturally the same exponents but reverses signs
for $B$ and $C$ staying signs of $A,E$ the same.
This calculation shows that the system have to possess by one orbit entering
the equilibrium $O$ as $s\to \infty$ and one orbit entering $O$ as $s\to -\infty.$
Both these orbits belong to the half-space $r>0$ since $A>0$ in both
cases. In particular, the dependence $u_2$ in $r$ is of power type $u_2\sim r^{7/4}.$
This is seen in the Fig.~\ref{fig:5}a where a homoclinic orbit found is shown
(soliton). It is clearly seen the tangency at the entrance to the
equilibrium. The related homoclinic orbits for $0<\lambda<1$ enter and
leave the equilibrium as $s\to \pm\infty$ with different asymptotics that
can be seen on the Fig.~\ref{fig:9}a. This becomes clear below.

Sometimes our simulations performed with the initial equation (\ref{4ord})
not with the system. Then, to find symmetric homoclinic and periodic
solution we sought for orbits which intersect either one or two times the fixed point
set $Fix(L).$ For initial differential equation (\ref{4ord}) this plane
defines by relations $S'=0, S^{\prime\prime\prime}=0.$ The 3-plane $S' =0$ is a
cross-section for the majority of its points, so zeroes of the graph
$S^{\prime\prime\prime}(D^2)$ give values of $D^2$ for which homoclinic
orbits exist. As an example, such plot is presented on Fig.~\ref{fig:5}b.

As we shall see, the orbit behavior of the system reminds that for the case of
a saddle-center equilibrium
in a Hamiltonian system in two degrees of freedom \cite{L,LK,GR,MHR} (see details
below). Again, for small positive $D$ system (\ref{fin0}) is slow-fast with slow
variables $(r,v_1)$ and fast variables $(u_2,v_2).$ In this case it is informative to
investigate slow and fast systems \cite{GL} separately. Slow system is derived if
we set $D=0$ in the third and fourth equations, then we have either $r=0$ or
$v_2=0, u_2 = - v_1$. 3-plane $r=0$ divides the phase space into two half-spaces
$r>0$ and $r<0$. We shall work in the half-space $r>0$, here we get slow 2-plane
$v_2=0, u_2 = - v_1$. On this plane the slow system is given as
$$\dot r = -v_1,\;\dot v_1 = -r^2 + r^3/2.$$ Its structure is plotted on
Fig.2.

The fast systems is derived, if one changes $s\to s/D =\tau$, as a result
small parameter $D$ arises as a multiplier in the first two equations.
Then one sets $D=0$ that makes variables $r,v_1$ be parameters. The
third and fourth equation with parameters $r^0,v^0_1$ have integral
$h = (u_2+v_1^0)^2 + v_2^2$. The periods $2\pi/r_0$ of these linear systems
depend on $r^0.$ The motions in the full system, as $D>0$ small, in some thin
neighborhood of the slow manifold is a combination of two motions: the slow motion
along the orbits of the slow system and fast rotation around the slow orbit.
This follows from results of \cite{GL}. In particular, if one moves along the homoclinic
orbit of the slow system, then the orbit behavior looks very similar to that which is
observed near a homoclinic orbit to a saddle-center in a Hamiltonian system with two
degrees of freedom \cite{L,LK}.

\section{$\lambda \ne 0$: equilibria}

Now we turn to the case $D>0$ small enough.
One of our main concern is to find soliton solutions to the equation
(\ref{ini}). This corresponds to homoclinic orbits for equilibrium that
exists on the upper sheet (see below). In fact, there are two equilibria
on this sheet but only one of them has outgoing and ingoing orbits
(separatrices). The situation under consideration depends heavily on the
value of parameter $\lambda$ and changes at the ends of the segment $\lambda \in [0,1]$.

For positive $\lambda$ in the half-space $u_1 > -\lambda^2/2$ both systems on
4-dimensional sheets $V>-\lambda$ and $V<-\lambda$ are analytic Hamiltonian ones
\begin{equation}\label{sheets}
u_1' = u_2,\;v_1' = u_1+\lambda \mp\sqrt{\lambda^2 + 2u_1},\;Du_2' = v_2,\;
Dv_2' = -u_2 - v_1,
\end{equation}
with related Hamiltonians
$$
H=u_2v_1-\frac{u_1^2}{2}-\lambda u_1+\frac{u_2^2+v_2^2}{2}\pm\frac{1}{3}
(\lambda^2+2u_1)^{3/2}.
$$
So, all available methods can be applied to the study, in particular, it concerns
existence of homoclinic orbits and nearby dynamics. The equilibrium at the origin
$O(0,0,0,0)$ on the upper sheet for $0 < \lambda < 1$ is a saddle-center,
its eigenvalues are a pair of pure imaginary numbers and two reals. Indeed,
linearizing at $O$ gives a linear Hamiltonian system, its characteristic polynomial is
$$
D^2 \sigma^4 + \sigma^2 - (1-\lambda)/\lambda
$$
with roots
$$
\pm iD^{-1}\sqrt{(1+\sqrt{1+4D^2(1-\lambda)/\lambda})/2},\;\pm \sqrt{\frac{2(1-\lambda)}
{\lambda+\sqrt{\lambda^2+4D^2(1-\lambda)\lambda}}}.
$$
Coordinates of the second equilibrium on the upper sheet are $(2(1-\lambda),0,0,0)$, its
characteristic polynomial is
$$
D^2\sigma^4 + \sigma^2 + (1-\lambda)/(2-\lambda)
$$
with pure imaginary roots (the elliptic point)
$$
\sigma_1^2 = \frac{-1-\sqrt{1-4D^2(1-\lambda)/(2-\lambda)}}{2D^2},\;
\sigma_2^2 = -\frac{2(1-\lambda)/(2-\lambda)}{1+\sqrt{1-4D^2(1-\lambda)/(2-\lambda)}}.
$$
As $\lambda$ approach to $1-0$, both equilibria coalesce into one equilibrium
with non semi-simple double zero and two pure imaginary eigenvalues. The lower sheet
does not contain equilibria at all.

\subsection{Positive $\lambda > 1$}

For $\lambda > 1$ two equilibria exist on the upper sheet, $O$
and another one $P_+ = (2(1-\lambda),0,0,0)$, the latter exists for $1<\lambda <
2.$ Their characteristic equations are
$$
D^2\sigma^4+\sigma^2 + \frac{\lambda - 1}{\lambda}=0,\;
D^2\sigma^4+\sigma^2 + \frac{1-\lambda}{2-\lambda}=0.
$$
On the lower sheet there is an only equilibrium $P_- = (2(1-\lambda),0,0,0)$
which exists for $\lambda > 2$ with the characteristic equation
$D^2\sigma^4+\sigma^2 + \frac{\lambda-1}{\lambda-2}=0$.

The types of these equilibria depend on the value of $D^2$. These are as
follows. For the upper sheet we have
\begin{enumerate}
\item if $0<D^2<1/4$, then $O$ is an elliptic point;
\item if $D^2 > 1/4$, then $O$ is an elliptic point for $1<\lambda <
\lambda_0$, $\lambda_0 = 4D^2/(4D^2 -1)$, and $O$ is a saddle-focus for $\lambda
> \lambda_0;$
\item $P_+$ exists for $1<\lambda < 2$, within this interval it is a
saddle-center.
\end{enumerate}
For the lower sheet at $P_-$
\begin{enumerate}
\item $P_-$ exists for $\lambda > 2$, within this interval: if $0<D^2<1/4$ it is
a saddle-focus for $1<\lambda < \lambda_0$, $\lambda_0 = 1 + 1/(1-4D^2),$
and is an elliptic point for $2>\lambda > \lambda_0$;
\item if $D^2 > 1/4$, then $P_-$ is a saddle-focus for $\lambda > 2.$
\end{enumerate}

\subsection{Negative $\lambda$}

Negative values of the parameter $\lambda$ also have physical sense. Let
us first investigate the type of the equilibrium at $O$. Here we have
another distribution of equilibria on sheets in comparison with $\lambda >
0.$

On the upper sheet we have the only equilibrium $S_+ =(2(1-\lambda),0,0,0)$ and
on the lower sheet the unique equilibrium is $O =(0,0,0,0)$. Their types are
as follows. The characteristic equation for $S_+$ is $D^2\sigma^4 +
\sigma^2 + (1-\lambda)/(2-\lambda)=0$ and for $O$ is $D^2\sigma^4 +
\sigma^2 - (1-\lambda)/\lambda=0$.
For $S_+$ we get the following.
\begin{enumerate}
\item If $0<D^2<1/4$, then $S_+$ is an elliptic point for any $\lambda < 0;$
\item if $1/4<D^2<1/2$, then $S_+$ is an elliptic point for $\lambda_+ < \lambda < 0$
and is a saddle-focus point for $-\infty <\lambda < \lambda_+,$ $\lambda_+ =
-2\frac{2D^2-1}{1-4D^2}$;
\item if $D^2 > 1/2$, then $S_+$ is a saddle-focus point for any $\lambda < 0.$
\end{enumerate}
For $O$ we have
\begin{enumerate}
\item If $0<D^2<1/4$, then $O$ is a saddle-focus for $\lambda_- <\lambda <
0$, $\lambda_- = 4D^2/(4D^2-1)$, and $O$ is an elliptic point
for $-\infty <\lambda < \lambda_-;$
\item if $D^2>1/4$, then $O$ is a saddle-focus for any $\lambda < 0$.
\end{enumerate}

Suppose some solution of the upper system hits the branching plane
$u_1 = -\lambda^2/2$ at a finite value $y_0$ of ``time'' $y$ and its
tangent vector at this point is directed outward, i.e. to the half-space
$u_1 < -\lambda^2/2$, that is, $\dot{u}_1(y_0) = u_2(y_0) < 0$. To continue this
solution by means of the lower system we apply to the point
$m_+ = (-\lambda^2/2,v_1(y_0),u_2(y_0),v_2(y_0))$ the involution
$$
m_- = L(-\lambda^2/2,v_1(y_0),u_2(y_0),v_2(y_0)) =
(-\lambda^2/2,-v_1(y_0),-u_2(y_0),v_2(y_0)).
$$
At this symmetric point coordinates of $\dot{u}_1$ and $\dot{v}_2$ of the vector field
change signs but coordinates of $\dot{v}_1$ and $\dot{u}_2$ are the same. The
lower vector field on the 3-plane $u_1 = -\lambda^2/2$ coincides with the
upper vector field. Now we proceed the orbit from the point $m_-$
using the lower vector field for $y > y_0$, this trajectory enters to the
half-space $u_1 > -\lambda^2/2$. If its continuation reach again the
branching plane, we do the same using upper vector field. As was said above,
the value of $S$ does not change at these switchings but $S'$
and $S^{\prime\prime\prime}$ do. In this way we can get cavitons being
non-smooth homoclinic orbits which represent orbits joining stable and
unstable separatrices of the saddle-center and crossing 3-plane $u_1=
-\lambda^2/2$ under their journey.

In what follows, we perform simulations with the smooth system
(\ref{smooth}). If some solution to this system stay all time in the
half-space $r>0$, then this solution corresponds to the upper sheet
system. In particular, solitons correspond to homoclinic orbits of the
equilibrium $O (\lambda,0,0,0).$ Homoclinic orbits to $O$, which spend part
time in the half-space $r<0$, correspond to cavitons.

\section{Solitons and cavitons: homoclinic loops of saddle-center}

The existence of homoclinic loops to a saddle-center is a rather delicate
problem, since one needs to find the merge of one-dimensional stable and
unstable manifolds of the saddle-center within 3-dimensional singular
level of the Hamiltonian. This level is singular (it is not a smooth manifold
at any its point) because this level has a cone-type singularity at the
equilibrium. Thus, such a problem should be studied in a two-parameter unfolding
generally. The task becomes easier, if one considers reversible Hamiltonian
systems and searches for symmetric homoclinic orbits. Then generally an
unfolding has to be one-parametric (in fact, this depends on the type of an action
of the reversible involution near a saddle-center, \cite{L,MHR}). If we investigate
2-parameter families of reversible Hamiltonian systems, then one expects a possibility
to construct curves in the parameter plane along which the systems has homoclinic
orbits to the related saddle-centers. Remind that existence of a saddle-center for
Hamiltonian systems is a structurally stable phenomenon.

Since we are of interest with spectra on parameter $\lambda$, for which the
system has homoclinic orbits to the saddle-center, we recall the result
proved first in \cite{MHR}. There a general one-parameter unfolding of reversible
Hamiltonian systems was studied under an assumption that it unfolds a system with
a homoclinic orbit to a saddle-center and for this orbit a genericity condition
holds found first in \cite{L}. Then it was proved that the set of parameter
values which correspond to systems with symmetric homoclinic orbits to the
saddle-center (not obligatory 1-round ones) is self-limiting and self-similar:
each point is an accumulation point for this set. It is worth noting that
our system can have both solitons and cavitons. On the mathematical language
this corresponds to the case when both unstable
separatrices of the saddle-center can merge with stable ones forming one or even
two homoclinic loops. In case of one loop this can be impossible, if one deals
with the case B as was discovered in \cite{L} and was indicated in \cite{MHR}.

The orbit structure of an analytic Hamiltonian system near a homoclinic
orbit to a saddle-center was studied first in \cite{L,LK1} and then this was
extended to different situations including reversible systems
\cite{LK,MHR,GR,GR2,Yag,GLT}. The study is based on the reduction of
instead of studying the flow to the investigation of Poincar\'e
map and its orbit structure generated by the flow on some cross-section to a
homoclinic orbit. This is heavily facilitated by the usage of a local normal
form near a saddle-center due to Moser \cite{Moser}: there exists an analytic
symplectic local coordinates $(x_1,y_1,x_2,y_2)$ such that the Hamiltonian
$H$ in these coordinates casts in the form $H = h(\xi, \eta),$
where $h$ is an analytic function in variables $\xi = x_1y_1,$ $\eta = (x_2^2 +
y_2^2)/2$, $h=\sigma \xi + \omega\eta +\cdots,$ $\sigma\omega\ne 0.$ Such normal
form is integrable, local functions $\xi, \eta$, as functions in
$(x_1,y_1,x_2,y_2)$, are local integrals of the flow generated by Hamiltonian $H$.
This easily allows one to construct local map from a cross-section to stable
separatrix to a cross-section to unstable separatrix. This map has a singularity
at the trace of of the stable separatrix but can be redefined to get a continuous
map everywhere and analytic at all points except for the trace of the separatrix.
Orbits of the system correspond to orbits of Poincar\'e map. so its studying gives
a complete information concerning orbit behavior of the flow. Principal elements of
this picture were found in \cite{L,MHR,GR,GR2}. In particular, suppose a homoclinic
orbit to the saddle-center exist and some genericity condition holds for it, then
each Lyapunov periodic orbit possesses within its level of $H$ four transverse
homoclinic Poincar\'e orbits \cite{L} implying the existence of complicated (chaotic)
dynamics nearby \cite{Smale, Shilnikov}.

Separatrices of the saddle-center are orbits (different from the equilibrium itself)
on two invariant analytic curves through the equilibrium, in Moser coordinates they
are $x_1=x_2=y_2=0$ and $y_1=x_2=y_2=0$ (strong stable $W^s$ and strong unstable
$W^u$ local manifolds). A two-dimensional center manifold near the saddle-center
is given as $x_1=y_1=0$, it is filled with Lyapunov saddle periodic  orbits lying
each in its own level of the Hamiltonian. These periodic orbits are saddle ones
in the related level of $H$.

\begin{figure}[ht]
	\centering
	\includegraphics[width=0.5\linewidth]{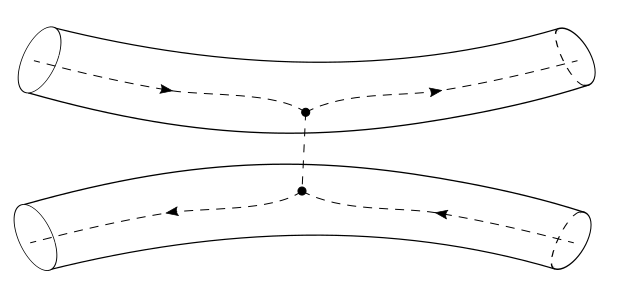}
	\caption{$c=0$, observe two glued points -- saddle-center}
\label{fig:6}
\end{figure}

\begin{figure}[ht]
\centering
\includegraphics[width=0.5\linewidth]{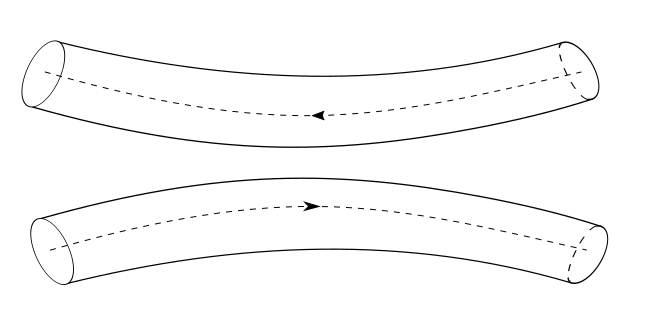}
\caption{$c<0$}
\label{fig:7}
\end{figure}

\begin{figure}[ht]
	\centering
	\includegraphics[width=0.5\linewidth]{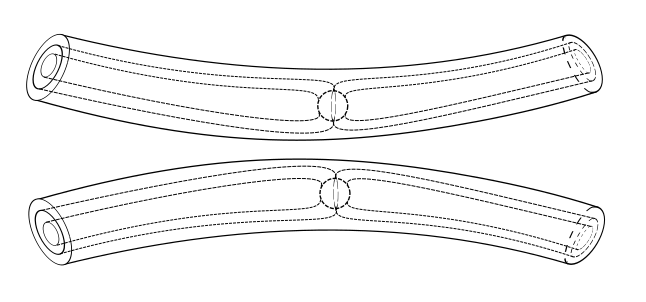}
	\caption{$c>0$}
\label{fig:8}
\end{figure}

The continuation of an unstable separatrix by the flow within the singular level
can lead to its merge with one of two stable separatrices forming a homoclinic orbit
to the saddle-center. The local orbit structure of the flow near such
orbit is rather well known since \cite{L} (see also \cite{LK,GR,MHR,GR,GR2,Yag}).
The orbit behavior depends essentially on the case which is realized of two
possible ones here \cite{L}. To explain this, let us remind the local structure
of the Hamiltonian near a saddle-center (see, Fig.~\ref{fig:6}-\ref{fig:8}). We
present here only related pictures (see details in \cite{LUt}). On these
pictures it is seen the local behavior of orbits, as well.

\begin{figure}[!ht]
\begin{minipage}[h]{0.4\linewidth}
\center{\includegraphics[width=1\linewidth]{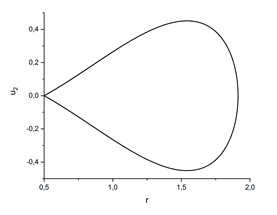} \\ a) }
\end{minipage}
\hfill
\begin{minipage}[ht]{0.4\linewidth}
\center{\includegraphics[width=1\linewidth]{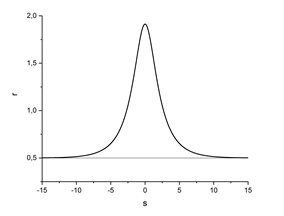} \\ b) }
\end{minipage}
\caption{{\footnotesize (a) 1-round homoclinic, $\lambda=0.5$, $D^2=0.23$; (b)
Unfolding of this 1-round soliton.}}
\label{fig:9}
\end{figure}

The cases mentioned depend on how the homoclinic orbit connects
cutting disks of two solid cylinders: the orbit can connect disks from the
same solid cylinder (case 1) or two different ones (case 2). As simulations
show that we deal with the case 2 for the system under study. Thus, two separatrices
going to the half-space $r>0$ may form homoclinic orbits corresponding
to solitons, and two remaining going to the half-space $r<0$ may form homoclinic
orbits corresponding to cavitons. Related orbits have been found, as an example,
they are plotted in Fig.\ref{fig:9}a,b. They represent both solitons and
solitons.

If the genericity condition mentioned above holds for a homoclinic orbit of the
saddle-center, then in the level $H=0$ containing the equilibrium and
the homoclinic orbit there exist also countably many saddle long periodic orbits
accumulating at the homoclinic orbit to the saddle-center \cite{L,LK}. As an example,
such periodic orbit is shown on Fig.~\ref{fig:17} but its fact there are many of them.
For the system we study all this picture takes place at fixed values of parameters
$D_*,\lambda_*$ for which a homoclinic orbit to a saddle-center exists.

When varying parameters $D,\lambda$, the orbit structure of the flow varies.
In particular, a homoclinic orbit to the saddle-center generically fails
to exist (it is destroyed). Instead, multi-round homoclinic orbits to $O$ can
arise \cite{Kolt}. Because saddle periodic orbits accumulate to the former
homoclinic loop, a situation may occur, when an orbit
on an one-dimensional unstable manifold of the saddle-center (which persists under
small changes $D,\lambda$) gets lie on the stable manifold of some saddle periodic orbit
$\gamma$ in the same level of $H$. Since the system under consideration is, in addition,
reversible, and if saddle-center $O$ and saddle periodic orbit $\gamma$ are symmetric,
then pairing orbit of the stable manifold of the saddle-center gets lie by symmetry on
the unstable manifold of $\gamma$. Thus, in this case a heteroclinic connection is
made up of two heteroclinic orbits, a symmetric saddle-center and a symmetric
saddle periodic orbit $\gamma$.

\begin{figure}[ht]
	\centering
	\includegraphics[width=0.5\linewidth]{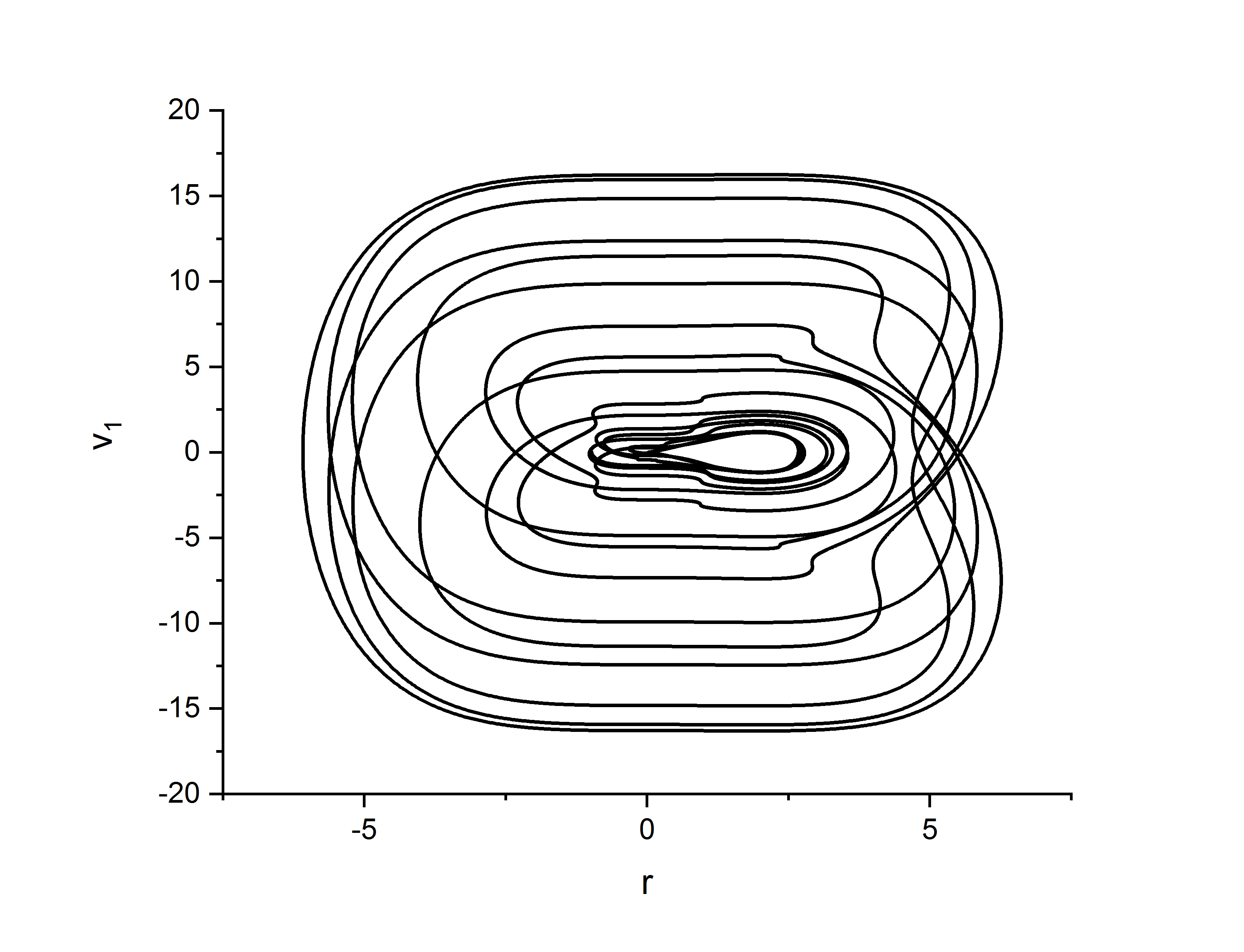}
	\caption{Heteroclinic connection in the singular level of $H$.}
\label{fig:10}
\end{figure}

Such heteroclinic connection can be of two types in dependence of how two
heteroclinic orbits are displaced with respect to two local solid cylinders
$H=H(p)$, described above. Namely, they can either intersect both the same
cylinder (case 1) or one heteroclinic orbit intersects one cylinder, but
another one does another cylinder (case 2). Our simulations show that we
deal here with the case 2. This implies that orbits leaving the unstable
manifold of $\gamma$ can return to lie on its stable manifold only making
at least one passage near two remaining stable and unstable manifolds of
the saddle-center. For our case this means that such orbits have to
intersect the plane $r=0$ before they return to the stable manifold of
$\gamma$. Existence of a heteroclinic connection of the type indicated is shown in
the Fig.\ref{fig:10}. Studying an orbit behavior near this connection was
performed recently \cite{LerTrif}.

Another feature of the system under varying parameters $(D,\lambda)$
is the appearance of new homoclinic orbits to the saddle-center.
Due to reversibility of the systems and the type of action of the reversor $L$
locally (the intersection of the fixed point set of $L$ with the singular level
$H=0$ is the curve through $p$) both homoclinic orbits (solitons) and
cavitons are usually symmetric orbits (i.e. invariant w.r.t. $L$). So, at a fixed
$D^2, \lambda$, $0<\lambda < 1$, only one symmetric soliton can exist and only one
symmetric caviton. Under varying parameters these homoclinic orbits usually are
destroyed, but can exist multi-round homoclinic orbits which before
closing make several excursions near the former 1-round
homoclinic orbits. Moreover, for a reversible system in the plane of
parameters $(D^2,\lambda)$ there are usually countably many bifurcation
curves accumulating to the curve of 1-round homoclinic orbits
\cite{MHR,GR2}. Our calculations show just this behavior, see,
Figs~\ref{fig:12}-\ref{fig:16}.

\begin{figure}[!ht]
\begin{minipage}[h]{0.4\linewidth}
\center{\includegraphics[width=0.9\linewidth]{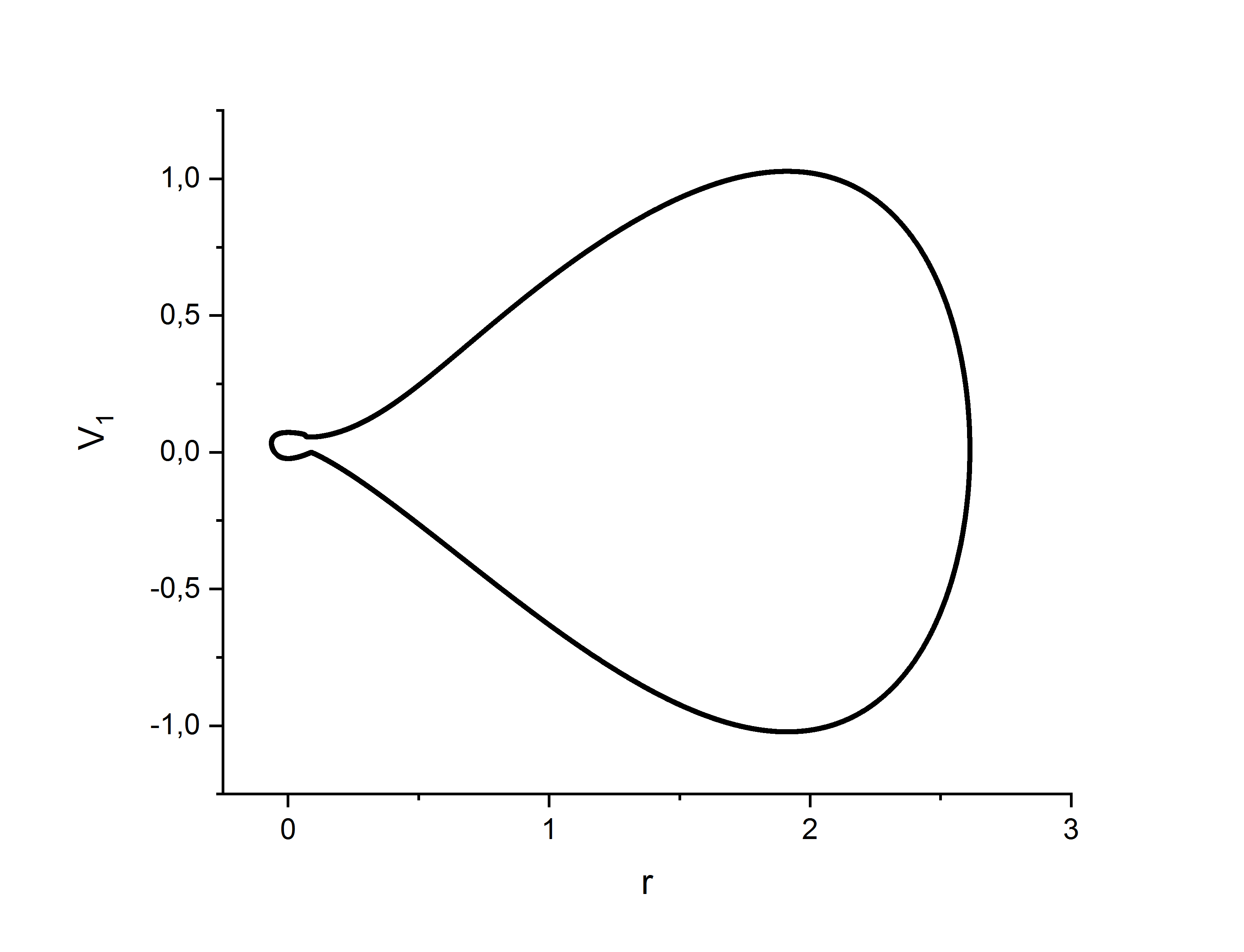} \\ a) }
\end{minipage}
\hfill
\begin{minipage}[ht]{0.4\linewidth}
\center{\includegraphics[width=1\linewidth]{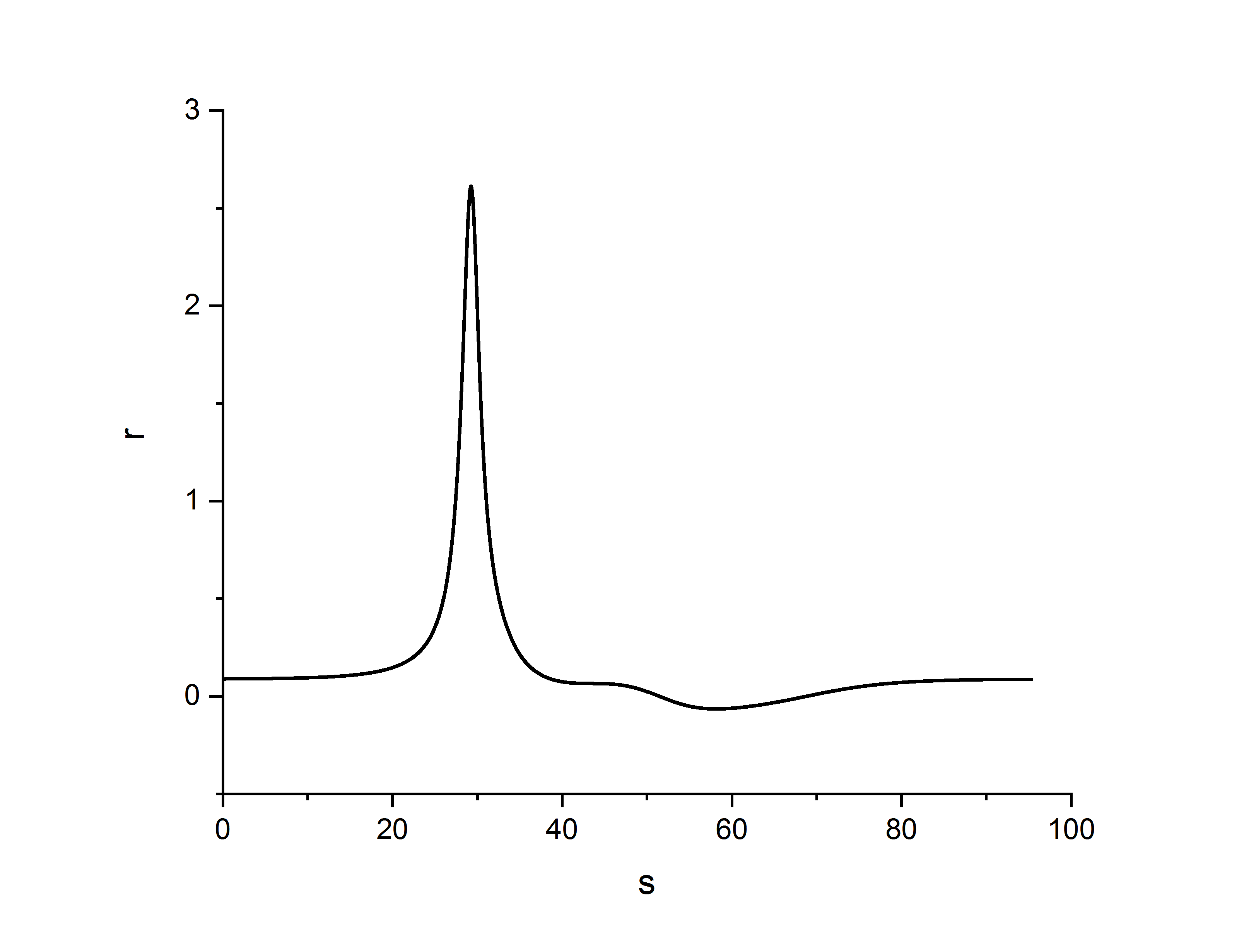} \\ b) }
\end{minipage}
\caption{{\footnotesize (a) Nonsymmetric soliton at ($\lambda=0.0887472, D^2=0.247906$); (b)
Its unfolding.}}
\label{fig:11}
\end{figure}

\begin{figure}[!ht]
\begin{minipage}[h]{0.4\linewidth}
\center{\includegraphics[width=1\linewidth]{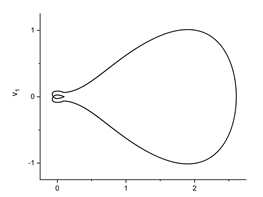} \\ a) }
\end{minipage}
\hfill
\begin{minipage}[ht]{0.4\linewidth}
\center{\includegraphics[width=1\linewidth]{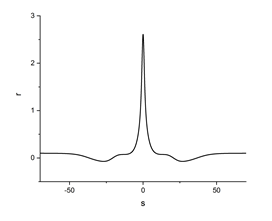} \\ b) }
\end{minipage}
\caption{{\footnotesize (a) Simplest caviton; (b)
Its unfolding.}}
\label{fig:12}
\end{figure}

\begin{figure}[!ht]
\begin{minipage}[ht]{0.4\linewidth}
\center{\includegraphics[width=1\linewidth]{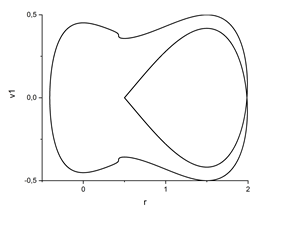} \\ a) }
\end{minipage}
\hfill
\begin{minipage}[h]{0.4\linewidth}
\center{\includegraphics[width=1\linewidth]{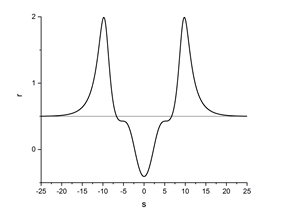} \\ b) }
\end{minipage}
\caption{{\footnotesize (a) 2-round homoclinic orbit, $\lambda =0.5, D^2=0.47$; (b)
Its unfolding -- 2-hump soliton.}}
\label{fig:13}
\end{figure}

\begin{figure}[!ht]
\begin{minipage}[ht]{0.4\linewidth}
\center{\includegraphics[width=1\linewidth]{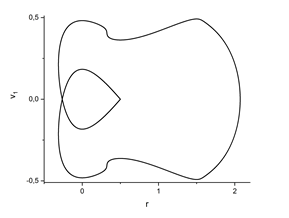} \\ a) }
\end{minipage}
\hfill
\begin{minipage}[ht]{0.4\linewidth}
\center{\includegraphics[width=1\linewidth]{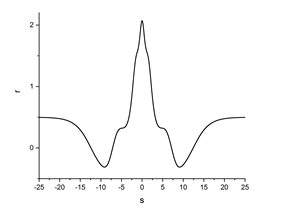} \\ b) }
\end{minipage}
\caption{{\footnotesize (a) 2-round homoclinic orbit with sharpening, $\lambda =0.5,
D^2=0.188847$; (b) Its unfolding -- 2-hump caviton.}}
\label{fig:14}
\end{figure}

\begin{figure}[!ht]
\begin{minipage}[ht]{0.4\linewidth}
\center{\includegraphics[width=1\linewidth]{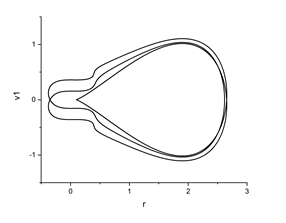} \\ a) }
\end{minipage}
\hfill
\begin{minipage}[ht]{0.4\linewidth}
\center{\includegraphics[width=1\linewidth]{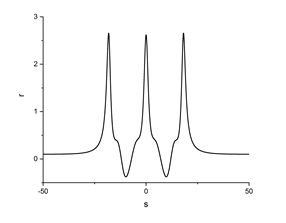} \\ b) }
\end{minipage}
\caption{{\footnotesize (a) 3-round homoclinic orbit, $\lambda =0.1, D^2=0.315485$; (b)
Its unfolding -- 3-hump soliton.}}
\label{fig:15}
\end{figure}

\begin{figure}[!ht]
\begin{minipage}[ht]{0.4\linewidth}
\center{\includegraphics[width=1\linewidth]{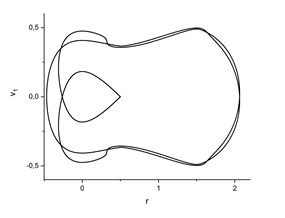} \\ a) }
\end{minipage}
\hfill
\begin{minipage}[ht]{0.4\linewidth}
\center{\includegraphics[width=1\linewidth]{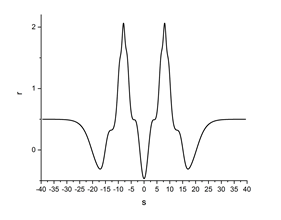} \\ b) }
\end{minipage}
\caption{{\footnotesize (a) 3-round homoclinic orbit with sharpening, $\lambda =0.5,
D^2=0.181595$; (b) Its unfolding -- 3-hump caviton.}}
\label{fig:16}
\end{figure}

\begin{figure}[ht]
	\centering
	\includegraphics[width=0.5\linewidth]{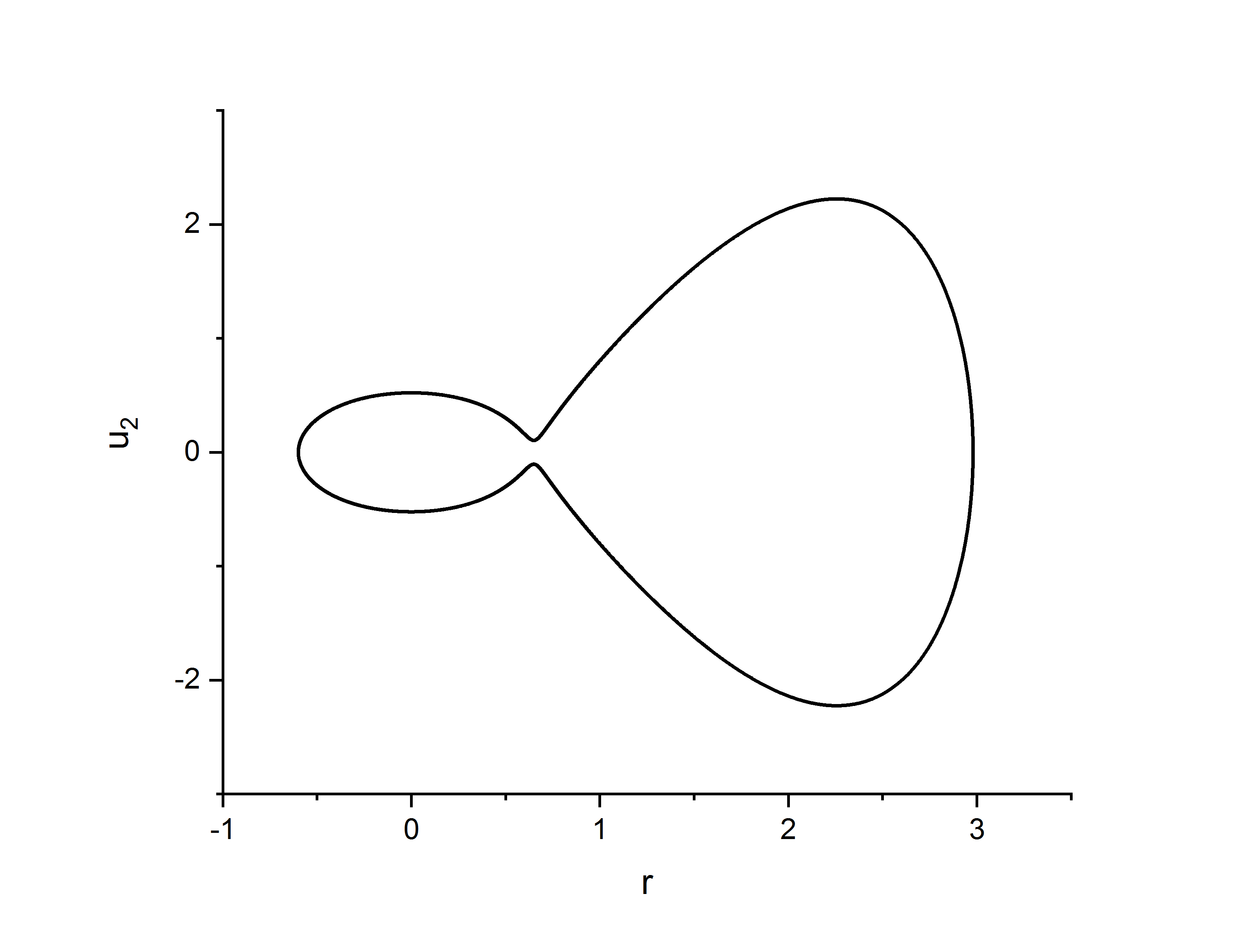}
	\caption{Saddle periodic orbit in the singular level of $H$.}
\label{fig:17}
\end{figure}

One more situation that can arise under varying parameters
$(D^2,\lambda)$, is the existence of nonsymmetric homoclinic orbits. By
symmetry, if such an orbit exists, there is another homoclinic orbit being
the symmetric counterpart of the former. The existence of such
nonsymmetric homoclinic orbit requires the 2-parameter analysis, they
exist at selected points (see Fig.\ref{fig:11}) a,b.

\section{Small positive $1-\lambda$}

Let us now study the problem for small positive $1-\lambda$ near the point $(0,0,0,0)$
on the upper sheet. As was said above, this equilibrium is degenerate at $\lambda =1$
with double zero eigenvalue and two imaginary eigenvalues $\pm i\omega.$ Let us scale
the initial equation (\ref{ini}):
$$
\lambda = 1-\eps^2,\;\tau = \eps y,\;\frac{d}{dy}=\eps\frac{d}{d\tau},\;S=\eps^2 X,\;
D=\frac{\kappa}{\eps}.
$$
As a result of these transformations we come to the following equation
\begin{equation}\label{norm}
\displaystyle{\kappa^2\frac{d^4X}{d\tau^4}+ \frac{d^2 X}{d\tau^2}- X + \frac{1}{2}X^2 -
\eps^2 X (1-\frac{3}{2}X + \frac{1}{2}X^2)+ \eps^4(-X +3X^2 - \frac{5}{2}X^3 +
\frac{5}{8}X^4)+\cdots},
\end{equation}
that defines the behavior of solutions as $\lambda \to 1-0.$ As above, let
us reduce the equation to the Hamiltonian system by means of the change of
variables $u_1=X,\; u_2 = X',$ $v_1 = -X' - \kappa^2 X^{\prime\prime\prime},$
$v_2 = \kappa X^{\prime\prime}$. The equation is reduced to the slow-fast
Hamiltonian system with respect to the symplectic form
$dv_1\wedge du_1 + \kappa dv_2\wedge du_2$
\begin{equation}\label{scale}
\begin{array}{l}
u_1' = u_2,\;v_1' = -u_1+ \frac{1}{2}u_1^2 - \eps^2 u_1(1-\frac{3}{2}u_1 +
\frac{1}{2}u_1^2)+
\eps^4(-u_1 +3u_1^2 - \frac{5}{2}u_1^3 + \frac{5}{8}u_1^4)+\cdots,\\
\kappa u_2' = v_2,\; \kappa v_2' = -(u_2 + v_1).
\end{array}
\end{equation}
The Hamiltonian of the system is
$$
H= u_2v_1 + \displaystyle{\frac{1}{2}(u_2^2 + v_2^2) + \frac{1+O(\eps^2)}{2}u_1^2 -
\frac{1+O(\eps^2)}{6}u_1^3 + \frac{\eps^2(1+O(\eps^2))}{8}u_1^4 -
\frac{\eps^4(1+O(\eps^2))}{8}u_1^5 +\cdots}.
$$
In this form we get a problem about an orbit behavior near a ghost separatrix
on an almost invariant elliptic manifold where a saddle equilibrium with a homoclinic
orbit exists. Such problem was studied partially in \cite{Gel_Lerm_JMS}.

The system (\ref{scale}) is also reversible with respect to the involution
$L:(u_1,v_1,u_2,v_2)\to (u_1,-v_1,-u_2,v_2)$ with its fixed point set
$Fix(L) = \{v_1=0,\,u_2=0.\}$

In order to find a homoclinic orbit to the saddle-center it is important
to keep in mind that we have a reversible slow-fast Hamiltonian system whose
fast system is a fast rotation. Indeed, to get the fast system, we do the scaling
of the independent variable $\tau/\kappa = \xi$, then the small multiplier $\kappa$
appears in the right hand sides of the first and second differential
equations. Then, setting $\kappa = 0$ we get $u_1,v_1$ as parameters of the
system of two remaining equations. They are linear and have an equilibrium -- center --
on any leaf $u_1=u_1^0,\;v_1=v_1^0.$ Thus, all assumptions of the theorem
1 from \cite{GL} hold and therefore there is a neighborhood $U$ of a compact region
in the slow plane $u_2=-v_1,$ $v_2=0,$ where analytic Hamiltonian by an analytic
symplectic change of variables is transformed to the function $H = H_0 (I, u,v,\kappa) +
R(x,y,u,v,\kappa)$, $I = (x^2+y^2)/2$, $|R| = O(\eps[-c/\kappa]).$ Thus,
in $U$ the Hamiltonian is exponentially close to an integrable Hamiltonian
$H_0$ with $I$ being an additional integral. In particular, this theorem works
for a region $U$ which contains on the slow plane the separatrix loop of the saddle.
Also, in this case a theorem from \cite{Gel_Lerm_JMS} holds which asserts the validity
of the Moser normal form \cite{Moser} for $H$ in some neighborhood of saddle-center of
the size $O(C\eps).$ These two theorems allows two prove the following theorem being
an analog of the theorem 1 from \cite{AM}
\begin{theorem}
For a small positive $\kappa$ in the whole phase space a neighborhood of the order
$\kappa$ of the homoclinic orbit on the slow manifold exist such that two branches of
stable and unstable separatrices of the saddle-center which cut the cross-section
$x_2 =0$ first time are displaced on the distance of the order
$R\exp[-c/\kappa]$ with some positive constants $R,c.$
\end{theorem}

For $\eps =0$ the equation above is the well studied, it also models the form of
stationary water waves on the surface of a liquid with the surface tension
\cite{AK,HM,AM}, if we eliminate terms of the order $\eps^2$ and higher.
It was proved for small $\kappa$ \cite{AM} this equation to have not localized solutions,
or, in other terms, no homoclinic solutions to the corresponding saddle-center exist
for the related slow-fast Hamiltonian system. Nevertheless, our simulations have
shown the existence of homoclinic orbits under varying $\eps,\kappa$ (see,
as a hint, Fig.\ref{fig:5}b). More exactly, the following hypothesis seems to be
valid

{\bf Hypothesis}. There is a neighborhood of the point $(0,0)$ on the parameter plane
$(\kappa,\eps)$ such that a countable set of bifurcation curve exists
which correspond to the existence of homoclinic orbits of any roundness.

\section{Homoclinic loops to saddle-focus}

\begin{figure}[!ht]
\begin{minipage}[ht]{0.5\linewidth}
\center{\includegraphics[width=1\linewidth]{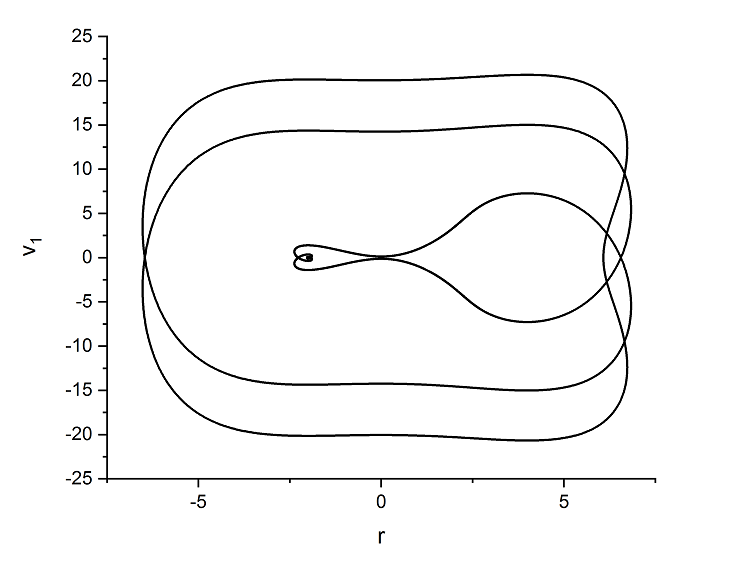} \\ a) }
\end{minipage}
\hfill
\begin{minipage}[ht]{0.5\linewidth}
\center{\includegraphics[width=1\linewidth]{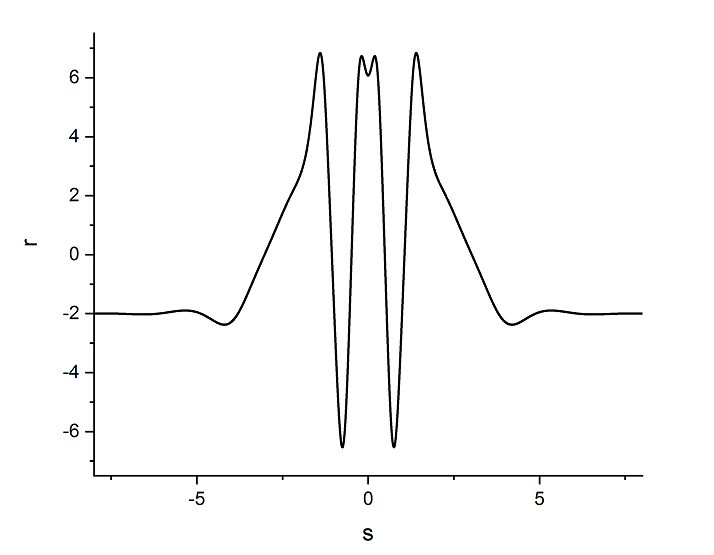} \\ b) }
\end{minipage}
\caption{{\footnotesize (a) Saddle-focus homoclinics, $D^2=0.36,\,\lambda=-2$;
(b) and its unfolding.}}
\label{fig:19}
\end{figure}

\begin{figure}[!ht]
\begin{minipage}[ht]{0.5\linewidth}
\center{\includegraphics[width=1\linewidth]{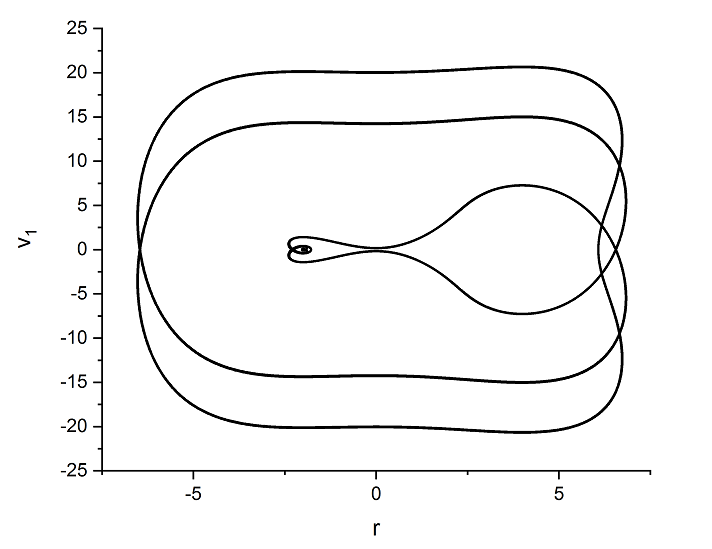} \\ a) }
\end{minipage}
\hfill
\begin{minipage}[ht]{0.5\linewidth}
\center{\includegraphics[width=1\linewidth]{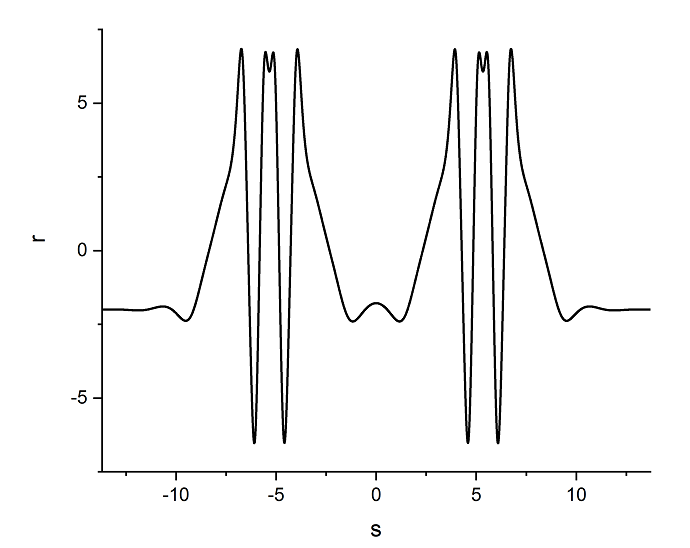} \\ b) }
\end{minipage}
\caption{{\footnotesize (a) Saddle-focus homoclinics, $D^2=0.36,\,\lambda=-2,
\mbox{\rm 2-rounded}$;
(b) and its unfolding.}}
\label{fig:20}
\end{figure}

\begin{figure}[!ht]
\begin{minipage}[ht]{0.5\linewidth}
\center{\includegraphics[width=1\linewidth]{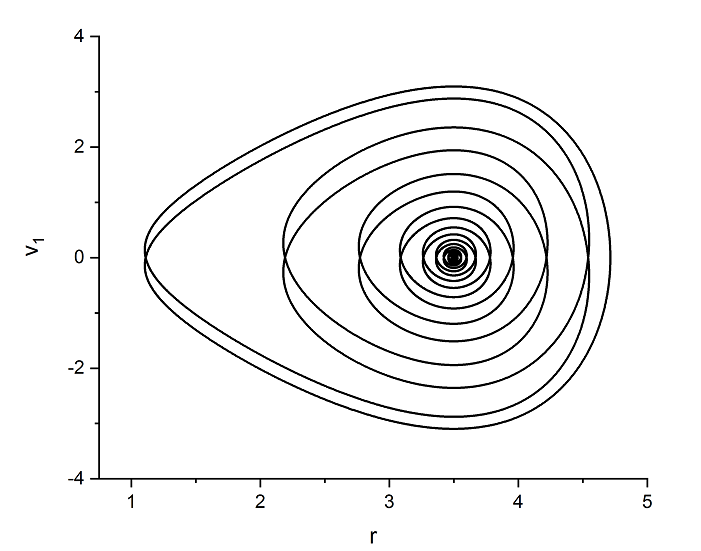} \\ a) }
\end{minipage}
\hfill
\begin{minipage}[ht]{0.5\linewidth}
\center{\includegraphics[width=1\linewidth]{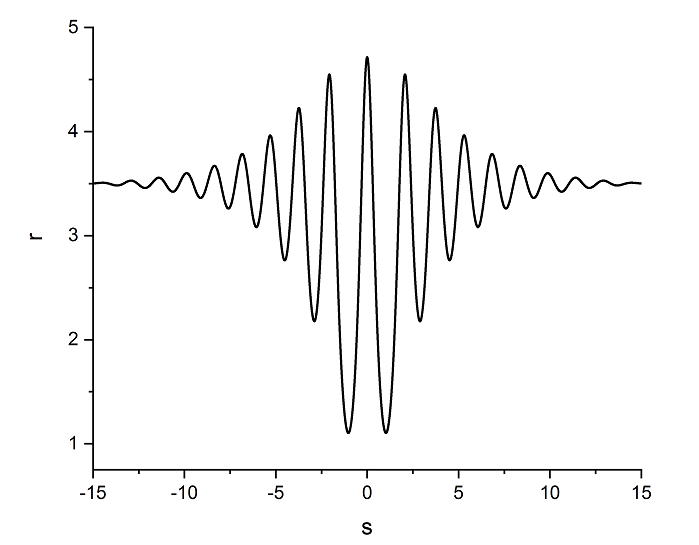} \\ b) }
\end{minipage}
\caption{{\footnotesize (a) Saddle-focus homoclinics, $D^2=0.36,\,\lambda=3.5,$\;
and (b) its unfolding.}}
\label{fig:21}
\end{figure}

\begin{figure}[!ht]
\begin{minipage}[ht]{0.5\linewidth}
\center{\includegraphics[width=1\linewidth]{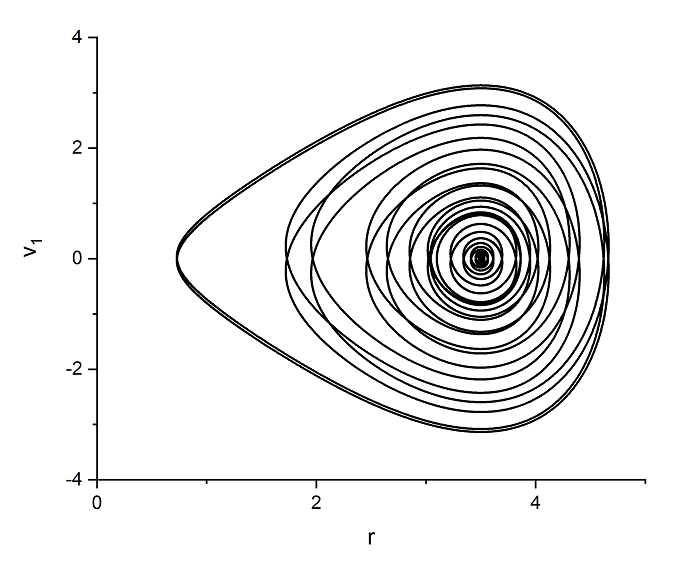} \\ a) }
\end{minipage}
\hfill
\begin{minipage}[ht]{0.5\linewidth}
\center{\includegraphics[width=1\linewidth]{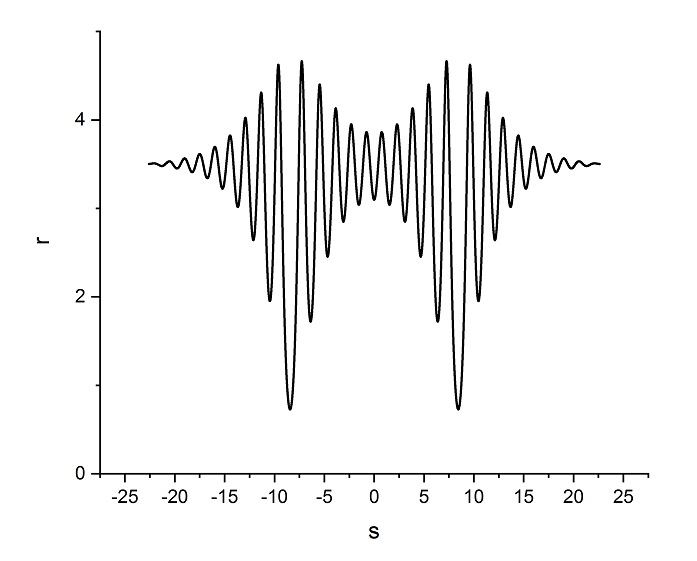} \\ b) }
\end{minipage}
\caption{{\footnotesize (a) Saddle-focus homoclinics, $D^2=0.36,\,\lambda=3.5,\;
\mbox{\rm 2-rounded}$
and (b) its unfolding.}}
\label{fig:22}
\end{figure}

The calculations of equilibria and their types show, in particular, that
if $D^2 > 1/4$, then for positive $\lambda > \lambda_0$ the equilibrium $O$ on the
upper sheet is the saddle-focus. The simulations discovered the abundance
of symmetric homoclinic orbits to this equilibrium (see Fig.\ref{fig:19}-\ref{fig:22}).
These homoclinic orbits are usually transverse in the following sense.
The related singular 3-dimensional level of the Hamiltonian (containing the
saddle-focus) includes both smooth 2-dimensional stable and unstable manifolds of the
saddle-focus and their intersection along the homoclinic orbit is
transverse within this level. The theory of the complicated orbit behavior near
a saddle-focus loop was developed by Shilnikov \cite{SHil} for general systems and
later adapted \cite{Dev} to cover the case of Hamiltonian systems (see also
\cite{Dev1,Hart,Champ1} where some elements of complex dynamics were proved for
reversible systems. An overview of these results can be found in \cite{HomSand}).
It says that near a transverse homoclinic orbit there exists multi-pulse homoclinic
orbits and a complicated behavior of nearby orbits (hyperbolic subsets) \cite{Dev}.
Moreover, varying levels of the Hamiltonian leads to many bifurcations of
hyperbolic sets, creations of elliptic periodic orbits, etc. \cite{Ler1,Ler2}.
Also, for the system under study there
are those symmetric homoclinic orbits which intersect during their travel
the branching plane $u_1=-\lambda^2/2$. Such homoclinic orbits can also be
named {\em cavitons with oscillating asymptotics} at infinity. Near them
multi-pulse cavitons also exist as well as a complicated orbit structure.

Let us prove the existence of two symmetric homoclinic orbits to the
equilibrium $O$ on the upper sheet for small enough $\lambda-\lambda_0.$
To do this, we use results of studying the Hamiltonian Hopf bifurcation
\cite{Meer} and their realization for the Swift-Hohenberg equation
\cite{GleLer,BK}. Recall that the Hamiltonian Hopf bifurcation is the
bifurcation in an one-parameter family of Hamiltonian systems in two degrees
of freedom having equilibria for all values of a parameter and at some critical value
of a parameter the related equilibrium has two double pure imaginary
eigenvalues each with the 2-dimensional Jordan box (non-semisimple case).
The type of bifurcation that occurs under transition through this critical
value of the parameter depends on the sign of some coefficient in the
normal form of the Hamiltoniian near this equilibrium. In particular, if
this coefficient is positive, then for those values of the parameter, when
the equilibrium is a saddle-focus, the system, if it is, in addition,
reversible, the saddle-focus gives the birth of two small symmetric homoclinic
orbits. The reversibility here guarantees their existence, otherwise, to
find such orbits is a very delicate problem related (for analytic systems)
with exponentially small splitting of stable and unstable manifolds of the
saddle-focus \cite{G}.

We prove the result reducing our problem to that being similar to the problem
as for the Swift-Hohenberg equation. To that end, let us scale the traveling
coordinate $y=\gamma \xi$, $\gamma = \sqrt{2}D,$ in the initial equation (\ref{4ord}).
After scaling and dividing at $D^2$ we get the equation
$$
u^{(IV)}+2u^{\prime\prime} + u = (1-4D^2 + \frac{4D^2}{\lambda})u - \frac{2D^2}
{\lambda^3}u^2 + \frac{2D^2}{\lambda^5}u^3 +\cdots.
$$
In notations of \cite{GleLer} we get $\alpha = 1-4D^2 +
\frac{4D^2}{\lambda},$ $\beta = - \frac{2D^2}{\lambda^3}$. The change $u\to
-u$ allows one to make $\beta$ positive as in \cite{GleLer}. Thus we get
the criterion of the birth of homoclinic orbit when crossing $\alpha = 0,$
i.e. we just get $\lambda = \lambda_0.$ Nevertheless, the equation differs from
\cite{GleLer}, since coefficient at the term with $u^3$ is positive. So,
we need to calculate the needed coefficient in the normal form directly.
We perform this calculation using the averaging. This was done long ago
\cite{norm} but unpublished.

We calculate the coefficient, we remark that saddle-foci appear as
$\lambda > \lambda_0$ as $D^2>1/4.$ Denote $-\nu = 1-4D^2 + \frac{4D^2}{\lambda}$
and consider $\nu$ as small positive parameter. After scaling $u = -\kappa
u$ with $\kappa = \sqrt{2}D/\lambda^{5/2}$ we come to the equation of the
form (we preserve old notations)
\begin{equation}\label{quasiSH}
u^{(IV)}+2u^{\prime\prime} + u = -\nu u + \beta u^2 + u^3 +\cdots.
\end{equation}
Let us write the equation in the form of two second order equations
$$
u^{\prime\prime}+u = v,\;v^{\prime\prime}+v = -\nu u + \beta u^2 + u^3 +\cdots.
$$
After scaling $u\to \sqrt{\nu},$ $v\to \nu v$ and denoting $\mu = \sqrt{\nu}$,
we get the system
$$
u^{\prime\prime}+u = \mu v,\;v^{\prime\prime}+v = \beta u^2 -\mu (u-u^3) + O(\mu^2).
$$
At $\mu =0$ we have the system whose solutions are of the form
$$
u=A\exp[i\xi]+\bar{A}\exp[-i\xi],\;v=B\exp[i\xi]+\bar{B}\exp[-i\xi]-
\frac{\beta}{3}(A^2\exp[2i\xi]+\bar{A}^2\exp[-2i\xi])+2\beta|A|^2.
$$
We add here new variables $u'=p, v'=q$ and differentiation of above
equalities under an assumption that $A,B$ are constant gives the relations
for $p,q$ through $A,\bar{A},B,\bar{B}$. We consider these relations as
the change of variables $(u,v,p,q)\to (A,\bar{A},B,\bar{B}).$ Observe that
this change of variables depend $2\pi$-periodically in $\xi.$

Performing this change of variables, we come to the system of four first order
differential equations in variables $(A,\bar{A},B,\bar{B})$ which is the
$2\pi$-periodic system in the so-called standard form of the averaging method (see,
\cite{BM}) $X' = \mu F(X,\xi)$. Averaging this system in $\xi$ gives the
average system $Y' = \mu F_0(Y)$,
$$
F_0(Y)= \frac{1}{2\pi}\int\limits_{0}^{2\pi}F(Y,\xi)d\xi.
$$
For our case we have
\begin{equation}\label{ave}
A' = -i\mu\frac{B}{2},\;B'=i\mu\frac{A}{2}[1-\frac{27+2\beta^2}{9}|A|^2],
\;\bar{A}' = c.c.,\;\bar{B}' = c.c.
\end{equation}
The coefficient we sought for is $\frac{27+2\beta^2}{9}$. It is positive
that means the existence of the homoclinic skirt in the system (\ref{ave})
which is integrable and the existence of two symmetric homoclinic orbits
in the initial system due to its reversibility \cite{IP}. The structure of
the averaged system is easily restored if introduce real variables
$(a,b,c,d),$ $A=a+ib,$ $B=c+id.$ In these variables we have a Hamiltonian
system
$$
a'=c,\;c'=\frac{a}{4}(1- L(a^2+b^2)),\;
b'=d,\;d'=\frac{b}{4}(1- L(a^2+b^2)),\;L= \frac{27+2\beta^2}{9}
$$
with Hamiltonian
$$H=\frac{c^2+d^2}{2}-\frac{a^2+b^2}{8}+\frac{L}{16}(a^2+b^2)^2$$
and an additional integral $K=ad-bc$. The common level $H=K=0$ gives the
homoclinic skirt, i.e. one-parameter family of homoclinic orbits to the
equilibrium $O$ of a saddle type with merged 2-dimensional stable and
unstable manifolds.

In the similar way one can check that zero equilibrium $O$ on the lower sheet
for negative $\lambda$ also gives the birth of homoclinic orbits to a
saddle-focus as $0<D^2<1/4$ at $\lambda=\lambda_-.$

As we have seen above, saddle-foci in the system exist both for negative
$\lambda$ and for positive $\lambda.$ Results of our simulations in these
cases are plotted in Fig.~\ref{fig:19},\ref{fig:20} for negative
$\lambda$. For positive $\lambda$ we get the following plots,
Fig.~\ref{fig:21},\ref{fig:22}.

\section{Conclusion}

In this work we have studied localized traveling wave solutions of the nonlocal
Whitham equation by means of the reduction to a Hamiltonian system. This initial
equation is of the fourth order with a nonlinearity being double-valued.
The reduction allows to derive a two degrees of freedom Hamiltonian system
but it defined on the two-sheeted space due to the type of nonlinearity.
In addition, the system is reversible with respect to some involution.
This permitted to obtain a clear geometric description of both smooth
solutions and solutions with singularities and apply to the problem of
developed methods of the theory of Hamiltonian dynamics, in particular,
theory of homo- and heteroclinic orbits. The search for homolinic and
heteroclinic orbits in dynamical systems is a very nontrivial problem,
being global in its own nature. Therefore, numerical methods with the
sharp set up allow to solve this problem for the concrete equation like that
under study. But the numerical search can be made much more rigorous if we
have some points in the parameter space (our $(D^2,\lambda)$) at which the
system has degenerate equilibria. Then bifurcation methods allows one to
find homoclinic orbits through the bifurcation. We do this using
Hamiltonian Hopf bifurcation and calculation the needed coefficients in the
local normal form to determine the type of the bifurcation. All this
together allowed one to investigate the system with many details.

\section{Acknowledgement}

The work by L.Lerman was partially supported by the Laboratory of Topological Methods
in Dynamics NRU HSE, of the Ministry of Science and Higher Education of RF, grant
\#075-15-2019-1931 and by Ministry of Science and Higher Education of Russian Federation
(Project \#0729-2020-0036). Numerical simulations were performed under a
support of the Russian Foundation of Basic Research (grants 18-29-10081, 19-01-00607).

\end{document}